\documentclass[
10pt,aps,amsmath,amssymb,
reprint,
pra
,longbibliography
,notitlepage
]{revtex4-1}
\usepackage{dcolumn}
\usepackage{bm}
\usepackage{braket}
\usepackage{xcolor}
\usepackage{graphicx}
\usepackage{mathtools}
\usepackage{multirow}
\usepackage[colorlinks=true,linkcolor=blue,urlcolor=black,citecolor=blue,bookmarksopen=true]{hyperref}
\usepackage{tabularx}
\usepackage[caption=false]{subfig}

\begin{document}

\title{Proposal for generating complex microwave graph states using superconducting circuits}
\author{Chenxu Liu}

\author{Edwin Barnes}

\author{Sophia Economou}
\affiliation{Department of Physics, Virginia Tech, Blacksburg, Virginia, 24061, USA}

\date{\today}

\begin{abstract}
Microwave photonic graph states provide a promising approach for robust quantum communication between remote superconducting chips using microwave photons. Recently, Besse \textit{et al}~\cite{Besse2020} demonstrated that 1D graph states can be generated using two transmon qubits. In this paper, we propose to use transmon qubits combined with other microwave devices to construct more complex graph states. Specifically, we consider 2D lattice and tree-like graph states. We compare the performance using fixed- versus tunable-frequency transmon qubits and also for different photonic qubit encodings. In each case, we estimate the fidelity of the resulting microwave graph state assuming current experimental parameters and identify the main factors that limit performance.
\end{abstract}

\maketitle

\section{Introduction}

Quantum computing is expected to achieve considerable speedup relative to certain classical computing algorithms. Examples include Grover's search algorithm and Shor's factorization algorithm~\cite{Grover1996,Shor1994,Shor1997,Simon1997}. Superconducting circuits are one of the most promising platforms for quantum computing owing to their relatively easy fabrication and fast controllability~\cite{Sheldon2016, Barends2013, Barends2014, Kjaergaard2020, DevoretReview2017, Girvin2011circuit}. In recent years, the number of transmon qubits that can be integrated into a single superconducting chip has increased significantly to beyond 50 qubits~\cite{Arute2019, Gambetta2020, Chow2021, JWPan2021ZCZ}.

However, building a large-scale quantum computer based on superconducting circuits will ultimately require coupling multiple chips together, because the relatively large footprint of these circuits limits the number of qubits that can fit on a single chip inside a dilution refrigerator.
There is therefore a need for robust quantum communication between chips. Communication between remote transmon qubits has been demonstrated using a common transmission line (coplanar waveguide)~\cite{Leung2019, Zhong2019, Chang2020, Zhong2021} or using emission and absorption of an itinerant single microwave photon pulse~\cite{Kurpiers2018, Ibarcq2018}. However, making these types of quantum communication schemes robust remains an outstanding challenge.

To address this problem, we can take inspiration from quantum communication networks based on optical photons. In these systems, encoding the quantum information into photonic graph states, e.g., repeater graph states~\cite{Azuma2015, Buterakos2017, Li2019} and tree graph states~\cite{Varnava2006, Varnava2008, Zhan2020, Borregaard2021}, can help achieve robust quantum communication. Specifically, in 2015, Azuma \textit{et al.} introduced the all-photonic quantum repeater state for robust long-range quantum communication~\cite{Azuma2015}. In superconducting circuit systems, microwave photonic graph states can be used to robustly communicate between chips~\cite{Kurpiers2018, Ibarcq2018, Leung2019, Zhong2019, Chang2020, Zhong2021}, and transduce to the optical domain for long-range quantum state transfer~\cite{Higginbotham2018, Mirhosseini2020}. However, how to efficiently and accurately generate microwave graph states is still unclear.

In 2020, Besse \textit{et al.} experimentally demonstrated how to use transmon qubits to generate entangled photonic qubits at microwave frequencies~\cite{Besse2020}. Specifically, they showed how to produce 1D photonic graph states. However, more complex graph states are required for robust quantum communication and other quantum computing tasks based on microwave photons. For example, repeater graph states can facilitate robust quantum communication~\cite{Azuma2015, Buterakos2017, Li2019}, and 2D or higher-dimensional lattice graph states can be used as universal resource states for measurement-based quantum computing~\cite{Raussendorf2001PRL, Raussendorf2003,  Nest2006, Briegel2009}. It is essential to investigate how to robustly generate these complex microwave graph states with superconducting qubits. 

In this work, we consider how to generate two types of microwave graph states: 2D lattice (cluster) states and all-photonic repeater states. Specifically, we address the following three questions: (1) What superconducting circuits can generate these states? (2) Which type of transmon qubit is better, fixed-frequency or tunable-frequency transmons? (3) What photonic qubit encodings can we use and which give better performance? We start by briefly surveying the two types of transmon qubits and four types of photonic qubit encodings. We then propose generation circuits for the two types of microwave graph states and estimate the resulting state fidelities. Assuming coherence times and gate fidelities from current experiments, we show that a $2 \times 8$ qubit cluster state can be generated with fidelity $0.8$ for both types of transmon qubits. Moreover, we show that it is possible to generate a $6$-branch repeater graph state using either fixed-frequency or tunable-frequency transmons with fidelity $0.87$.

This paper is organized as follows. In Section~\ref{sec:toolbox}, we survey the toolbox for generating microwave photonic graph states. Specifically, we review the generation circuit of Ref.~\cite{Besse2020} in Section~\ref{subsec:coupler}, and compare two types of transmon qubits in Section~\ref{subsec:transmon_types} as well as four types of photonic qubit encoding methods in~\ref{subsec:encoding}. In Section~\ref{sec:2D_cluster}, we present generation circuits for 2D cluster states and estimate the achievable size of the generated state, while in Section~\ref{sec:rgs}, we propose generation circuits for repeater graph states and estimate the resulting state fidelities. In Section~\ref{sec:logical}, we consider using superconducting qubits to generate logically encoded photonic qubits. In Section~\ref{sec:summary}, we summarize our main results.

\section{Toolbox for generating microwave graph states using superconducting qubits} \label{sec:toolbox}

We start by considering what tools are available to us for generating microwave graph states. First, we review the generation scheme for 1D graph states demonstrated in Ref.~\cite{Besse2020}. We argue that using tunable couplers can afford us higher controllability. We then compare fixed-frequency and tunable-frequency transmon qubits and summarize their pros and cons. In addition, we compare four types of photonic qubit encodings, namely Fock-basis, time-bin, frequency-bin, and two-rail encoding methods. We summarize their advantages and disadvantages in the context of graph state generation.

\subsection{Tunable couplers for enhanced controllability} \label{subsec:coupler}

\begin{figure}[h]
    \centering
    \includegraphics[width = 0.4 \textwidth]{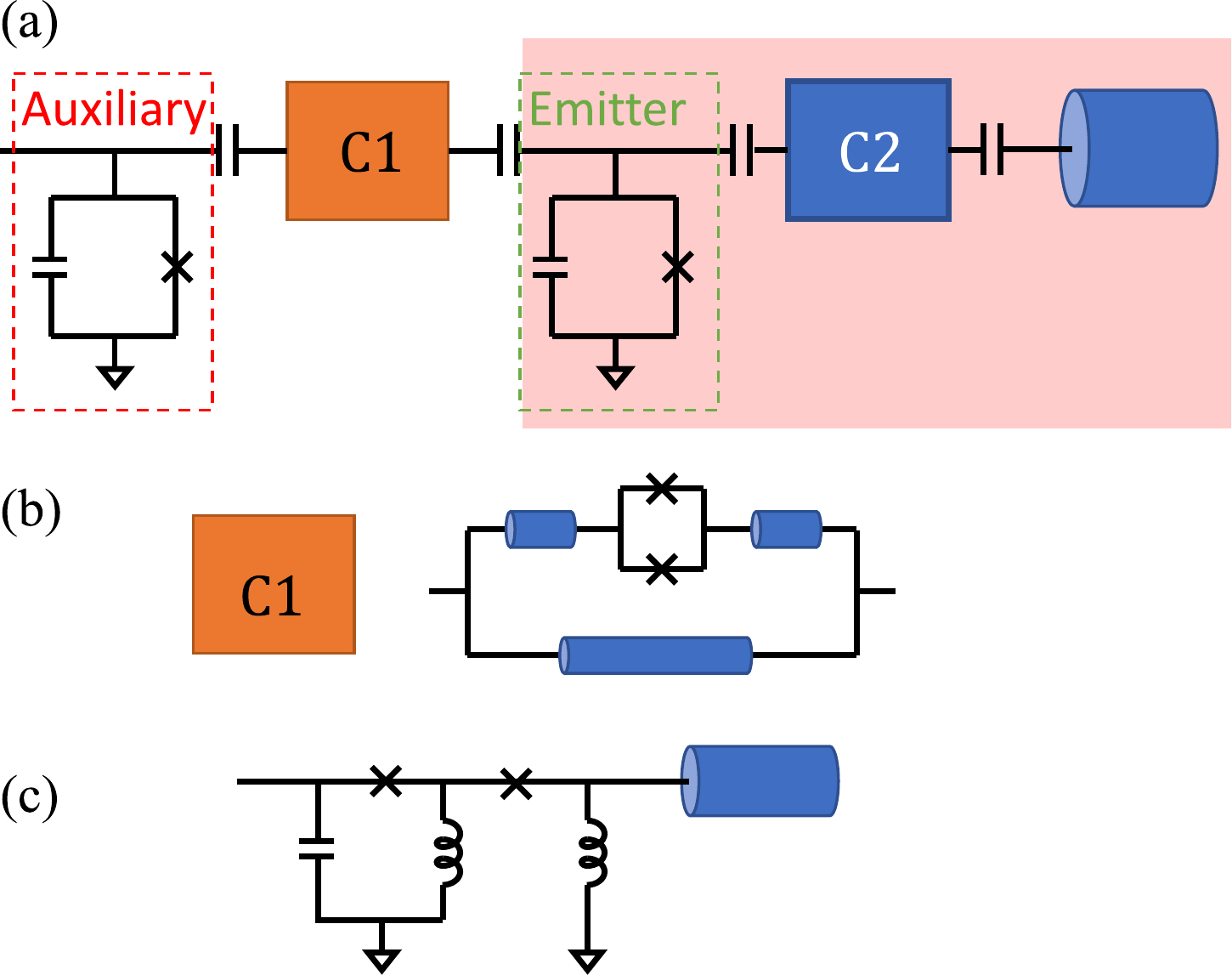}
    \caption{The circuit diagram for a modified microwave photon generation unit. In (a) we show the circuit representation of a microwave photon generation unit. There are two transmon qubits and a coupler to couple the two qubits together (labeled as C1). We add a second tunable coupler (labeled as C2) to connect the emitter qubit to the transmission line (labeled as C2). In (b) we show the coupler C1 that couples two transmons together. In (c), we show the circuit diagram corresponding to the shaded region in (a), which contains the tunable coupler C2.}
    \label{fig:coupler}
\end{figure}

In Ref.~\cite{Besse2020}, Besse \textit{et al}. experimentally demonstrated how to use two transmon qubits to deterministically generate 1D cluster states of microwave photons. Their approach was inspired by a prior scheme to create such states in the optical domain~\cite{Lindner2009}. In the optical case, the level structure of the quantum emitter is such that emitted photons are entangled with the emitter and with each other. The emission process can be described using a controlled-NOT (CNOT) operation between the emitter and the photonic qubit virtually initialized in the $\ket{0}$ state. This scheme does not directly apply to transmons since they do not possess the right level structure. However, as shown in Ref.~\cite{Besse2020}, this issue can be circumvented by using a second, auxiliary transmon. The CNOT is then implemented in two steps: (1) Apply a CNOT operation between the auxiliary qubit and the emitter qubit, and (2) the emitter qubit emits a microwave photon, which can be viewed as a SWAP operation between emitter and photon.

Note that in this microwave generation scheme, errors in the generated cluster states can come from (1) the decay and decoherence error of the auxiliary transmon qubit while it is waiting for the emitter to emit photons and (2) the imperfect gate fidelities on the transmon qubits. To reduce (1), we want to speed up the photon emission time, which requires reducing the lifetime ($T_1$) of the emitter qubit. On the other hand, to reduce the error from (2), we want to increase the lifetime and coherence time of the transmon qubits. To simultaneously address these seemingly conflicting requirements, we propose the use of a tunable coupler between the auxiliary qubit and the emitter qubit, as shown in Ref.~\cite{Besse2020}. This can help to reduce the static coupling between the two qubits, which can reduce the cross-talk error~\cite{Collodo2019, Mundada2019, Besse2020} (see Fig.~\ref{fig:coupler}a `C1'). To gain additional controllability and generate high-fidelity microwave graph states, we further propose to use another tunable coupler between the emitter qubit and the transmission line~\cite{Chen2014, Zhong2019, Zhong2021} (see Fig.~\ref{fig:coupler}a `C2').

In Fig.~\ref{fig:coupler}a, we show the proposed microwave photon generation unit (PGU). The PGU contains the two transmon qubits and the coupler C1 (shown explicitly in Fig.~\ref{fig:coupler}b) between them, as in Ref.~\cite{Besse2020}. The figure also shows the second tunable coupler C2 schematically, while a particular implementation of this coupler is shown in Fig.~\ref{fig:coupler}c. When the two-qubit CNOT gates are applied, the coupler C2 is turned off to increase the lifetime of the emitter qubit. When the emitter qubit needs to emit photons, the coupler C1 is turned off while C2 is turned on, which establishes strong coupling between the emitter qubit and transmission line modes. In Ref.~\cite{Zhong2019}, specifically, by tuning the tunable coupler between the transmon qubit and the transmission line, the decay rate of the transmon qubit to the transmission line ($\kappa/2\pi$) can be tuned from $\sim 0$ to $175$~MHz, where $\kappa \sim 0$ means the transmon qubit relaxation is dominated by its intrinsic loss, not the loss to the modes of the transmission line. When $\kappa/(2\pi) = 175$~MHz, the decay time constant is $\tau = 1/\kappa \sim 1.0$~ns. This greatly reduces the photon emission window while maintaining good two-qubit gate fidelity.

\subsection{Comparison of two types of transmon qubits} \label{subsec:transmon_types}

Transmons can be viewed as nonlinear LC oscillators, where the nonlinearity is provided by Josephson junctions. The nonlinearity makes the transition frequency from the first excited state to the second excited state detuned from the one between the ground state and the first excited state. This allows one to use the ground and first excited states as qubit levels. There are two types of transmon qubits: fixed-frequency (FF) transmons, and tunable-frequency (TF) transmons. In FF transmons, there is one Josephson junction to provide the nonlinearity, while a TF transmon has a DC SQUID loop instead of a single Josephson junction. This SQUID loop makes the frequency of the qubit tunable using a magnetic flux. 

\begin{table*}[t]
\caption{\label{tab:transmon_compare} Comparison of FF and TF transmons. We compare the native gate sets and the $T_1$, $T_2$ times for the two types of qubits. ``CR'' stands for cross-resonance gate.}
\begin{ruledtabular}
\begin{tabular}{l|ccll}
transmon type & Native 2Q gate & $T_1$, $T_2$ & Advantage & Disadvantage\\
\hline
\multirow{2}{*}{Fixed} & CR & $T_1 = 60~\mu$s ~\cite{Chang2013} & \multirow{2}{*}{long $T_1$ and $T_2$ time} & slow gate operation\\
& CNOT~\footnote{This is a CNOT-type gate, not exactly a CNOT gate. Specifically, the operation qubit is always initialized in the $\ket{0}$ state when the gate is applied.} & $T_2 = 55~\mu$s &  & slow photon emission\\
\hline
\multirow{2}{*}{Tunable} & \multirow{2}{*}{CZ} & $T_1 = 44~\mu$s ~\cite{Barends2014} & fast two-qubit gates & \multirow{2}{*}{short $T_1$ and $T_2$ time}\\
& & $T_2 = 20~\mu$s & fast photon emission & 
\end{tabular}
\end{ruledtabular}
\end{table*}

In Table~\ref{tab:transmon_compare}, we summarize the advantages and disadvantages of both FF and TF transmon qubits. As the TF transmon can be tuned by an applied magnetic flux, it is also more sensitive to flux noise compared to the FF transmon. Therefore, FF transmons can have longer lifetimes ($T_1 = 60~\mu$s) and coherence times ($T_2=55~\mu$s)~\cite{Chang2013} compared to TF transmons ($T_1 = 44~\mu$s, $T_2=20~\mu$s)~\cite{Barends2014}. However, TF qubits offer more controllability; for example, they allow for fast CZ gates ($\sim 40$~ns)~\cite{Barends2014}. In the case of FF transmons, the most widely used two-qubit gates are cross-resonance (CR) gates~\cite{Parauanu2006, Rigetti2010}. 
These are performed using microwave drives on both qubits to induce cross-resonance coupling. The resulting gate is slow ($\sim 200$~ns) compared to the CZ gate on TF transmons~\cite{Sheldon2016}. The fast gate operations on TF transmons can reduce the idling decay and decoherence error in the process of generating graph states.

\subsection{Comparison of four types of photonic qubit encodings} \label{subsec:encoding}

\begin{table*}[t]
\caption{\label{tab:encoding_compare} Comparison of Fock basis, time-bin, frequency-bin and two rail encodings of photonic qubits. We show the physical degrees of freedom (DOF) that encode the quantum information in each case. We also compare the advantages and disadvantages. `SQG' stands for single-qubit gate, while `TL' stands for transmission line.}
\begin{ruledtabular}
\begin{tabular}{l|lll}
Encoding & Physical DOF & Advantage & Disadvantage\\
\hline
\multirow{2}{*}{Fock-basis} & \multirow{2}{*}{presence/absence of a pulse} & Easy generation & SQGs challenging \\
            & & High fidelity to generate & low fidelity SQGs \\
\hline
\multirow{2}{*}{Time-bin} & \multirow{2}{*}{pulse in early/late time bins} & Can detect photon loss & Long generation time \\
            & & Easy to apply SQGs &  Low generation fidelity \\
\hline
\multirow{2}{*}{Frequency-bin} & \multirow{2}{*}{pulses with two frequencies}  & Can detect photon loss & SQGs require nonlinearity \\
            & & Faster generation~\footnote{Compared to time-bin encoding, frequency encoding can use parallelization to speed up. See main text for more details.} & Low generation fidelity\\ 
\hline
\multirow{3}{*}{Two-rail} & \multirow{3}{*}{pulses in two spatial modes} & Can detect photon loss & Low generation fidelity \\
            & & Faster generation~\footnote{Compared to time-bin encoding methods. Same as frequency-bin encoding.} & One qubit requires two TLs. \\
            & & Easier SQGs & 
\end{tabular}
\end{ruledtabular}
\end{table*}

There are multiple ways to encode quantum information into microwave photons. In this subsection, we survey four types of encodings: (1) Fock-basis, (2) time-bin, (3) frequency-bin, and (4) two-rail encoding. We summarize the four types in Table~\ref{tab:encoding_compare}, where we briefly compare the advantages and disadvantages of each encoding in generating microwave graph states. In the rest of the section, we discuss these encodings in more detail. Additional discussion on ways to generate the encoded photonic qubits is given in Appendix~\ref{appsec:photon_qubit_gene}.

Firstly, the Fock-basis encoded photon qubit uses the presence or absence of a single-photon pulse inside a fixed time window to encode 0 and 1 respectively. It only requires a single-photon emission window from a transmon qubit to generate a Fock basis qubit, as shown in Ref.~\cite{Besse2020}. Compared to the other three types of encoding, generating Fock-basis qubits is the fastest, which makes the generation scheme have higher fidelity. However, single-qubit gates (SQGs) on Fock-basis qubits require first absorbing the photon qubit into a transmon, which maps the photon state to the transmon state, and then applying the gate to the transmon and emitting the photon again~\cite{Reuer2021}. The gate fidelity of this capturing and re-emitting protocol is rather limited. Furthermore, Fock-basis qubits are especially susceptible to photon loss errors. If the photon gets lost, there is no way we can detect the error. 

Secondly, the time-bin encoding uses photon pulses that can appear in two possible time bins (early/late time-bin) to encode quantum information. Unlike the Fock-basis encoding, here SQGs can be applied using microwave photonic devices that are similar to optical beam splitters~\cite{Gu2017Review}, making them easier to implement. Furthermore, as there is always a single photon inside the two time-bins for each photonic qubit, the photon loss error can be detected. However, generating a time-bin qubit requires twice as much time as a Fock-basis qubit, which can lead to more error on the transmon qubits while they are waiting for the photon to be emitted.

Thirdly, the frequency-bin encoding method uses photon pulses with two possible frequencies to encode quantum information. Similar to the time-bin encoding, the frequency-bin qubit also allows for photon loss detection. Unlike time-bin qubits, however, here the photon pulses corresponding to the computational $\ket{0}$ or $\ket{1}$ states can overlap in the time domain, making it possible to parallelize the generation process (see Appendix~\ref{appsec:photon_qubit_gene}). Therefore, generating a single frequency-bin qubit requires less time compared to a time-bin qubit. However, as the photon pulses for the $\ket{0}$ and $\ket{1}$ states have two different frequencies, implementing SQGs requires nonlinear effects, which makes the gates more complicated compared to time-bin qubits. In the microwave regime, the coupling between two `colors' of microwave photons has been achieved using the nonlinearity provided by Josephson junctions~\cite{Zakka-Bajjani2011}. 

Lastly, the two-rail encoding uses photon pulses in two possible spatial modes to encode quantum information. Similar to the time-bin and frequency-bin encodings, it also allows for detecting photon loss errors. Compared to the time-bin encoding, the parallelization in generating a single two-rail qubit helps to speed up the generation process, similar to frequency-bin qubits. Furthermore, to apply SQGs on time-bin or frequency-bin qubits, it is necessary to map the pulses, either in the early or late time bins or with two possible frequencies, into two spatial modes that connect to the input ports of the microwave device that performs the gate operation. Such a mapping is unnecessary for two-rail encoded qubits. However, compared to the time-bin and frequency-bin qubits, one two-rail qubit requires two transmission lines, which may be experimentally expensive.

Comparing the four types of photonic qubit encodings, in order to generate a microwave graph state with the highest fidelity, the Fock-basis encoding is optimal. However, if SQGs on the photonic qubits are required, or if the photon loss error is significant and needs to be detectable, then the frequency-bin or two-rail encodings are better. If the on-chip microwave photonic devices can have efficient and high-fidelity frequency separation with little distortion on the input signal, the frequency-bin encoding is better as it has less experimental overhead. Otherwise, the two-rail encoding method is better in terms of achieving high fidelities in the generated graph states. In this rest of the paper, as microwave 2D cluster states and repeater graph states can be generated by applying gates solely on transmon qubits, we choose to use the Fock-basis encoding to achieve the highest state fidelities.

\section{Generating 2D microwave cluster states} \label{sec:2D_cluster}

In the previous section, we compared different types of transmon qubits and photonic qubit encodings. In the rest of the paper, we consider which of these options are more suitable for generating specific types of microwave graph states, and we present explicit protocols and devices for doing so. Specifically, we investigate two types of states: (1) 2D lattice graph (cluster) states, which are resource states for measurement-based quantum computing, and (2) repeater graph states, which are useful for robust quantum communication. In this section, we focus on generating 2D cluster states, while the generation of repeater graph states is discussed in Section~\ref{sec:rgs}.

To devise a method of creating 2D microwave cluster states from quantum emitters, we again take inspiration from results in the optical domain. In 2010, Economou \textit{et al.} proposed a scheme to generate optical 2D cluster states using quantum dots as photon emitters~\cite{Economou2010}. In this protocol, the 2D cluster state is generated by repeatedly applying CZ gates between emitters and pumping them to generate new columns of entangled photons. This approach combines the protocol of Ref.~\cite{Lindner2009} for generating 1D optical cluster states together with the principle that entangled emitters emit entangled photons to produce 2D cluster states. The CZ gates between the emitters create graph edges in the second dimension. 

\begin{figure}[h]
    \centering
    \includegraphics[width = 0.4 \textwidth]{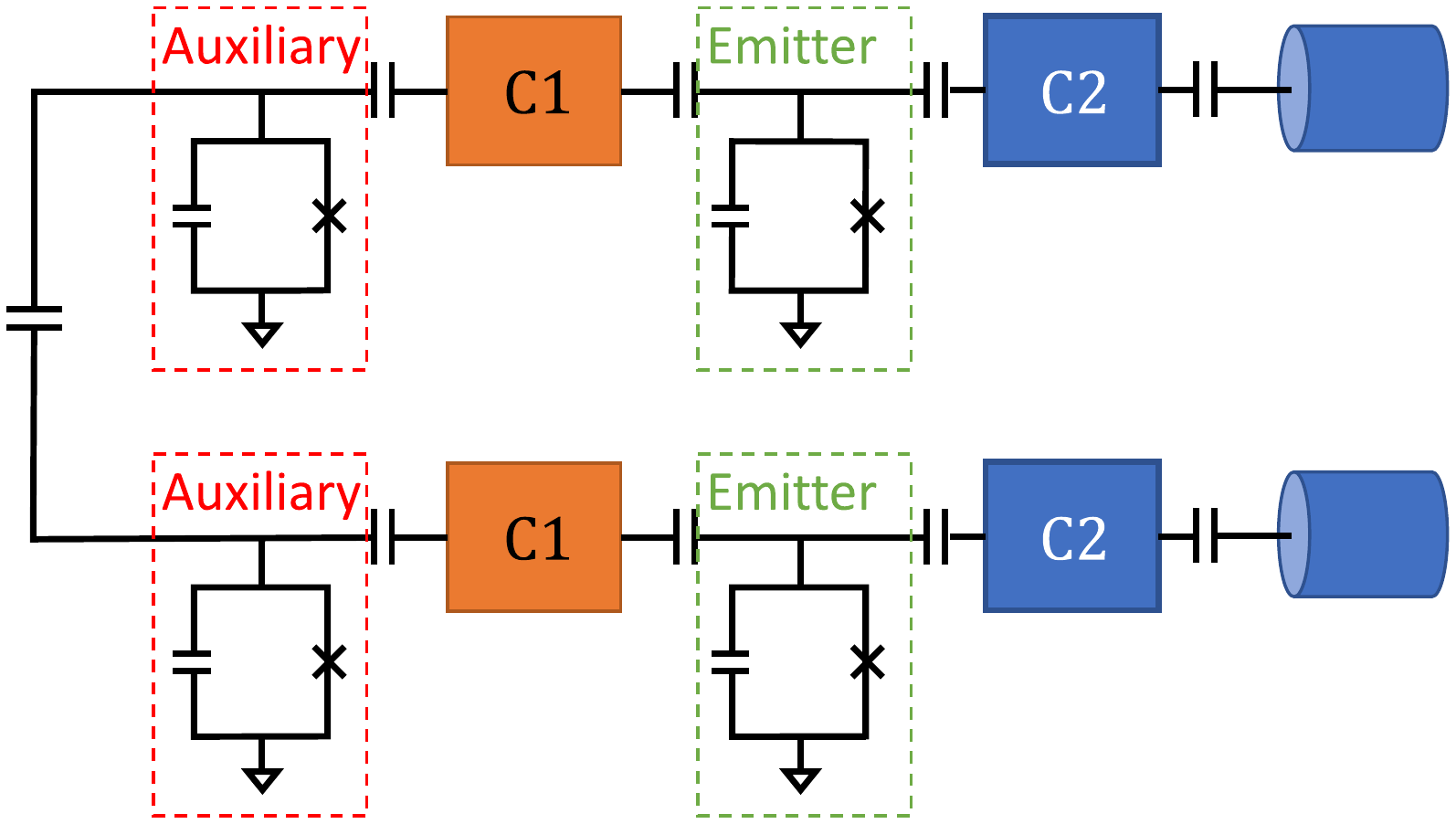}
    \caption{The generation circuit for creating 2D cluster states of microwave photons using transmon qubits.}
    \label{fig:2d_cluster_gene}
\end{figure}

Here, we adapt the central ideas of Ref.~\cite{Economou2010} to the microwave domain, where the quantum dots are replaced by transmon qubits. Fig.~\ref{fig:2d_cluster_gene} shows a schematic of a device capable of generating $2\times n$ microwave cluster states using two PGUs like those of Fig.~\ref{fig:coupler}. In this scheme, there is no need to apply SQGs on the photons in order to produce the target graph state, so we consider using the Fock-basis encoding for the photonic qubits. As before, each PGU contains two transmons, with one serving as the emitter and the other as the auxiliary qubit. The two auxiliary qubits from the two PGUs are capacitively connected, which enables the application of CZ gates between them. After initializing all the qubits in the two PGUs in the $\ket{0}$ state, the $2 \times n$ cluster state can be generated by repeating the following sequence of operations:
\begin{enumerate}
    \item Apply Hadamard gates on both auxiliary qubits in the two PGUs.
    \item Apply CNOT gates between the auxiliary qubit and the emitter qubit in each PGU.
    \item Apply a CZ gate between the two auxiliary qubits.
    \item Pump the emitter qubits to produce two photons.
\end{enumerate}
The full gate sequence for generating a $2 \times n$ cluster state can be found in Appendix~\ref{appsec:gate_2D_cluster}. A general 2D cluster state with $N = k \times n$ photons can be generated by adding another $k-2$ PGUs to the circuit in Fig.~\ref{fig:2d_cluster_gene} and repeating the above procedure $n$ times. In this case, the CZ gates need to be applied between all nearest-neighbor pairs of auxiliary qubits in adjacent PGUs. 

\begin{table*}[t]
\caption{\label{tab:2d_gates} Gate operation fidelities and coherence times for FF and TF transmons. $F_{\text{SQ}}$, $F_{\text{CZ}}$, $F_{\text{CR}}$, and $F_{\text{CNOT}}$ are the fidelities of SQGs, CZ gates, CR gates, and CNOT gates, respectively.
}
\begin{ruledtabular}
\begin{tabular}{l|c c c c|c c}
Transmon type  & $F_{\text{SQ}}$ & $F_{\text{CZ}}$ & $F_{\text{CR}}$  & $F_{\text{CNOT}}$& $T_1$ & $T_2$ \\
\hline
Fixed & $0.9995$~\cite{Barends2014} &  & $0.991$~\cite{Sheldon2016} & $0.928$~\cite{Besse2020} & $\sim60~\mu$s & $\sim 55~\mu$s~\cite{Chang2013}\\
Tunable & $0.9995$~\cite{Barends2014} & $0.995$~\cite{Barends2014} &  &  & $\sim 44~\mu$s & $\sim 20~\mu$s~\cite{Barends2013, Barends2014}
\end{tabular}
\end{ruledtabular}
\end{table*}

Next, we investigate the relative performance of FF and TF transmons in the generation circuit shown in Fig.~\ref{fig:2d_cluster_gene}. We do this by estimating the fidelity of the generated graph state. For each type of transmon, the two-qubit gates applied in the generation sequence need to be decomposed into native gate sets. For FF transmons, the CZ gates can be decomposed into SQGs and CR gates, while for TF transmons, the CZ gates are native, and hence they can be directly applied. In Table~\ref{tab:2d_gates}, we summarize the average fidelity of the single- and two-qubit gates for both types of transmons; below, we use these to estimate the fidelities of the resulting graph states.

We start from the FF transmon qubits. We consider using $k$ PGUs ($2k$ transmons) to generate a $k \times n$ microwave cluster state. Each generation cycle requires $4(k-1)$ SQGs, $k-1$ CR gates, and $k$ CNOT gates to generate a new column of photons. The average infidelity when generating a single column of photons is then
\begin{equation}
    \Delta F_{k}^{(\text{FF})} = 1 - F_{\text{SQ}}^{3(k-1)} F_{\text{CR}}^{k-1} F_{\text{CNOT}}^k F_{\text{idle}}^k,
    \label{eq:fix_2D_row}
\end{equation}
where $F_{\text{SQ}}$, $F_{\text{CR}}$, and $F_{\text{CNOT}}$ are the average fidelities of the SQGs, CR gates, and CNOT gates, respectively. $F_{\text{idle}}$ is the average idling fidelity of the auxiliary qubits in the presence of loss and decoherence during the photon emission window. Denoting this window by $\tau$, the idling fidelity is estimated to be 
\begin{equation}
    F_{\text{idle}} = \frac{1}{2} + \frac{1}{6}\left( e^{-\tau/T_1} + 2 e^{-\tau/T_2} \right),
    \label{eq:decay_decoh_loss}
\end{equation}
where $T_1$ and $T_2$ are the relaxation and coherence times of the qubits (see Appendix~\ref{appsec:idle_F}). To generate 2D cluster states of size $N = k \times n$, the emission cycle needs to be implemented $n$ times. So the infidelity of generating this $N$-photon-qubit cluster state is
\begin{equation}
    \Delta F_{N}^{(\text{FF})} = 1 - F_{\text{SQ}}^{3 N - 3 n} F_{\text{CR}}^{N-n} F_{\text{CNOT}}^N F_{\text{idle}}^N.
    \label{eq:fix_2D_all}
\end{equation}

Next, we consider using TF transmon qubits instead. As CZ gates between TF transmons can be performed quickly, we decompose the CNOT gates into CZ gates and Hadamard gates. Therefore, to generate the same $N = k \times n$ microwave cluster state from $k$ PGUs, each emission cycle has $3k$ SQGs and $2k-1$ CZ gates. The average infidelity when generating a single column of photonic qubits is approximately
\begin{equation}
    \Delta F_{k}^{(\text{TF})} = 1 - F_{\text{SQ}}^{3k} F_{\text{CZ}}^{2k-1} F_{\text{idle}}^{k}.
    \label{eq:tunable_2D_row}
\end{equation}
Repeating the generation cycle $n$ times, we can generate a $k \times n$ cluster state of microwave photons. The state fidelity is estimated to be
\begin{equation}
    \Delta F_{N}^{(\text{TF})} = 1 - F_{\text{SQ}}^{3N} F_{\text{CZ}}^{2N-n} F_{\text{idle}}^{N}.
    \label{eq:tunable_2D_all}
\end{equation}
We can understand the dependence on $F_{\text{idle}}$ as follows. First, notice that to generate an $N$-photon 2D cluster state, we always need $N$ CNOT gates or CZ gates between the auxiliary and emitter qubits. Furthermore, to generate a $k\times n$ microwave 2D cluster state using $k$ PGUs, the $k$ auxiliary qubits need to wait for $n$ photon generation time windows. So the total idling fidelity caused by decay and decoherence of the auxiliary qubits is $F_{\text{idle}}^{N}$, which is consistent with Eqs.~\eqref{eq:fix_2D_all} and~\eqref{eq:tunable_2D_all}.

Looking at Eqs.~\eqref{eq:fix_2D_all} and \eqref{eq:tunable_2D_all}, we notice that for a fixed total number $N$ of photons, the state fidelity is largest when the number of photon emission cycles $n$ is large, or equivalently, when we use a small number of PGUs. To understand this, first observe that the graph edges defining the 2D cluster state can be generated either by the photon emission from the PGUs or by applying the CZ gates between adjacent PGUs. When generating an edge via photon emission, the state fidelity drops by a factor $F_{\text{SQ}} F_{\text{emit}}$, where $F_{\text{emit}}$ is the average fidelity of the photon emission due to the finite emission window. As we assume the emission time window is at least $10$ times the emitter qubit lifetime, we have $F_{\text{emit}} > 99.99\%$. Therefore, the state fidelity drops by a factor $\sim F_{\text{SQ}}$. However, generating an edge in the 2D cluster state by applying CZ gates between the auxiliary qubits makes the state fidelity drop by a factor $F_{\text{CZ}}$. For FF transmons, we have $F_{\text{CZ}} \sim F_{\text{SQ}}^2 F_{\text{CR}} < F_{\text{SQ}}$, and similarly for TF transmons, we directly have $F_{\text{CZ}} < F_{\text{SQ}}$. Therefore, using fewer PGUs reduces the number of CZ gates relative to SQGs and thus improves the fidelity of the 2D cluster state. However, we stress that this consideration does not include the possible error caused by the loss of photons when they travel along the transmission line. When early-generated photonic qubits need to be stored in a delay line for further processing steps, photon loss becomes significant. Introducing more PGUs can counteract photon losses by reducing photon storage time requirements. The optimal number of PGUs will then be the one that balances CZ gate errors against photon loss errors. 

\begin{figure*}[htbp]
    \centering
     \subfloat[]{\includegraphics[width= 0.45 \textwidth]{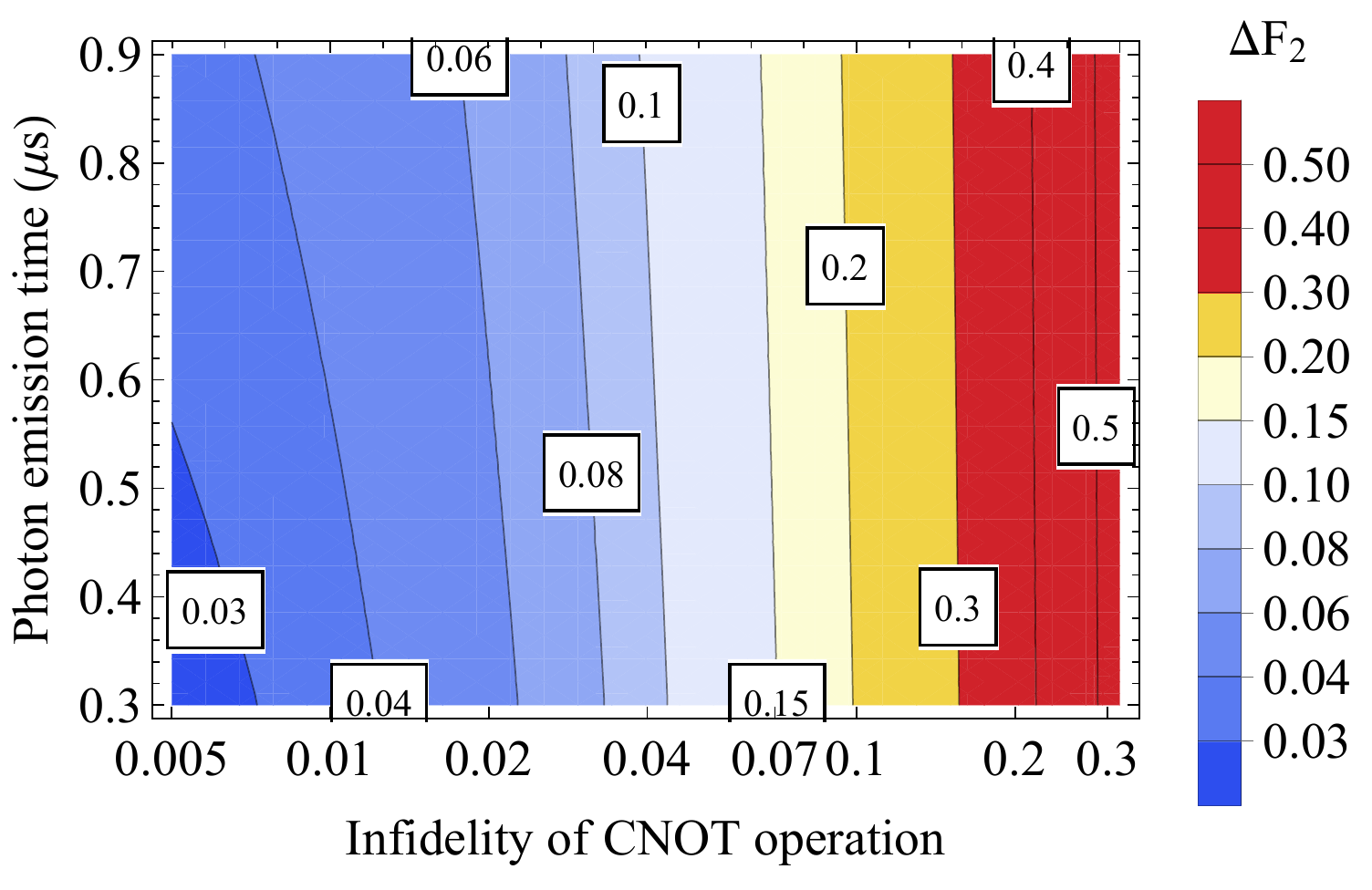}} \,
    \subfloat[]{\includegraphics[width= 0.45 \textwidth]{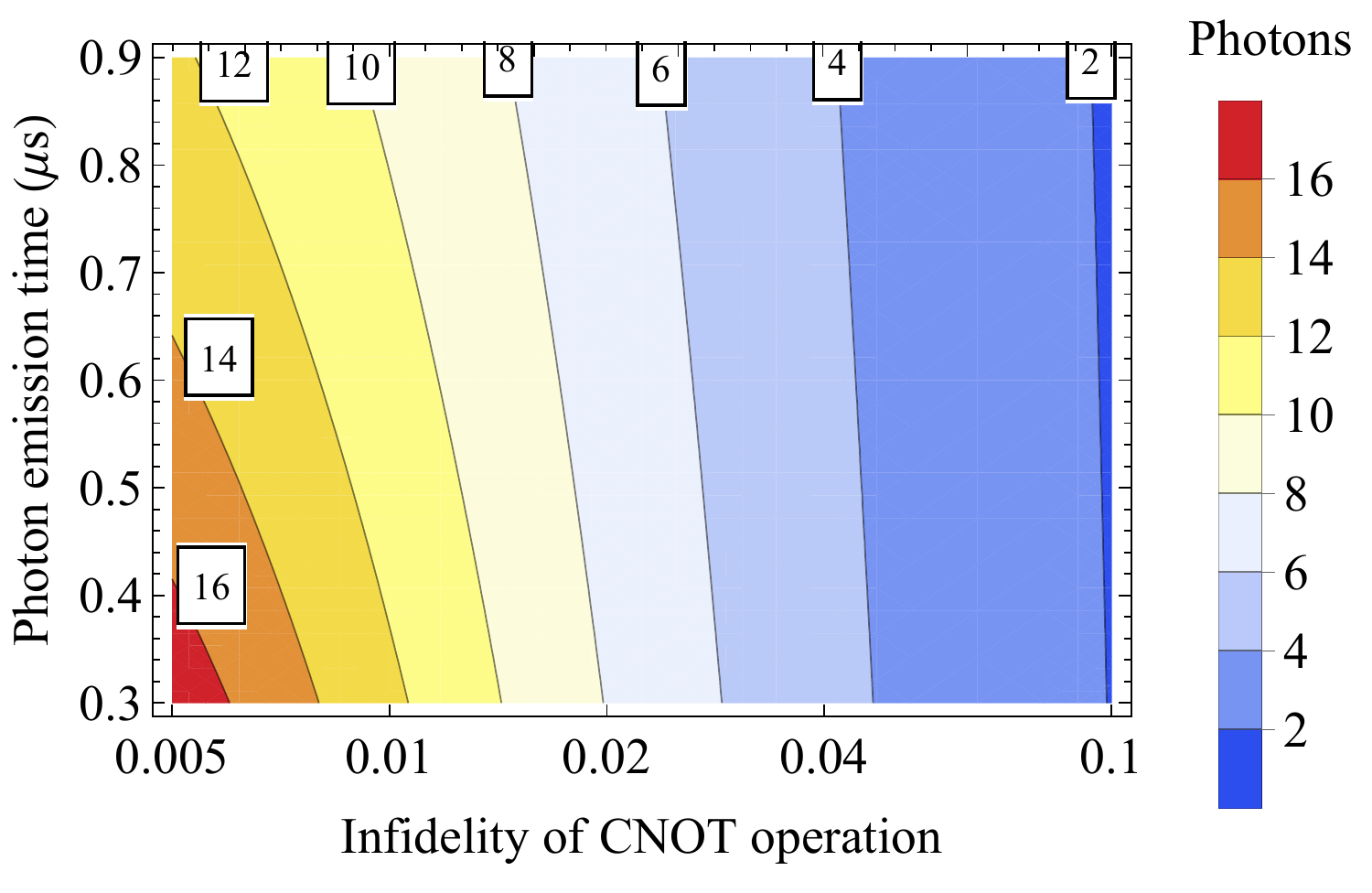}} \\
    \subfloat[]{\includegraphics[width= 0.45 \textwidth]{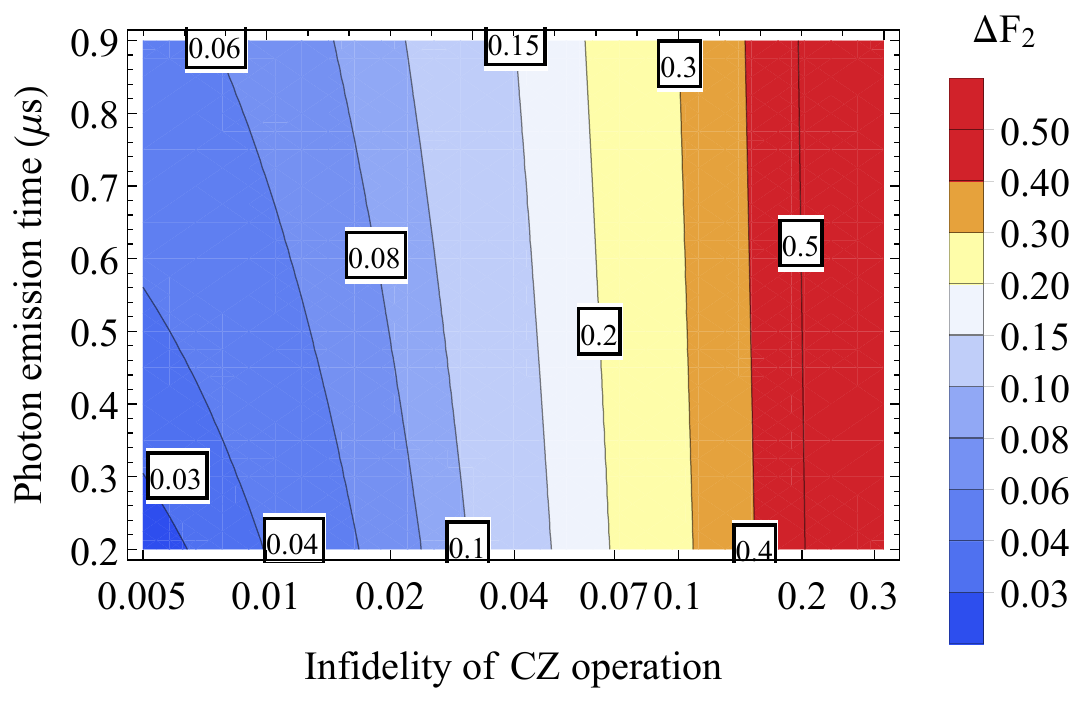}} \,
    \subfloat[]{\includegraphics[width= 0.45 \textwidth]{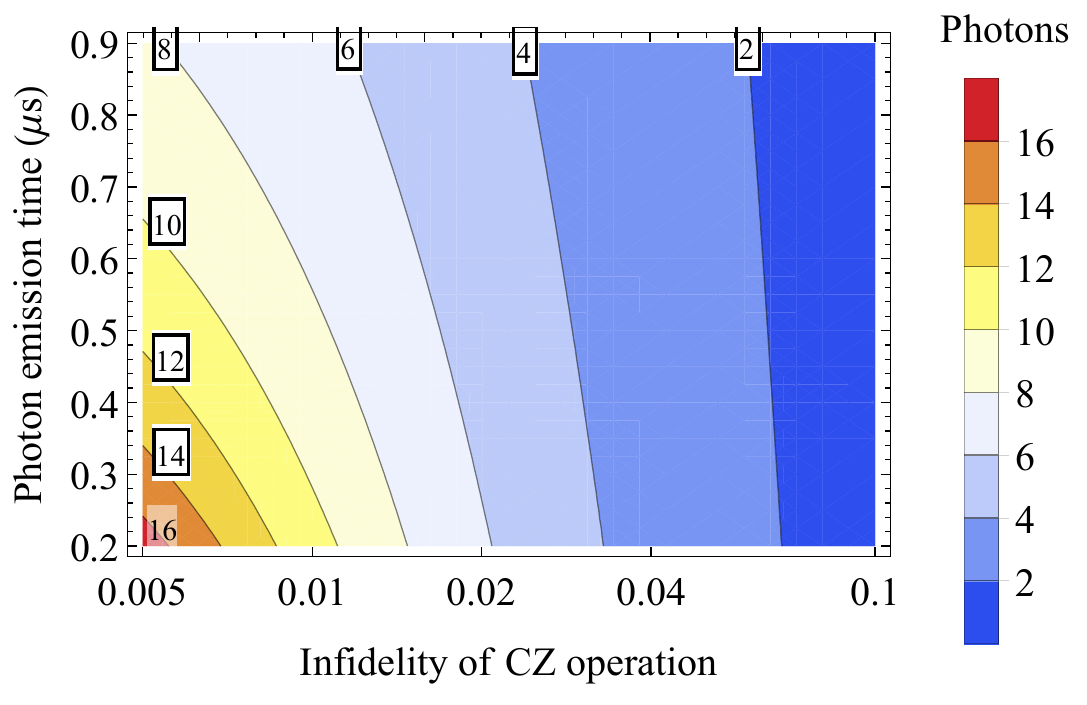}}
    \caption{(a,c) Infidelity in generating a single 2-photon column of a $2 \times n$ cluster state using (a) FF and (c) TF transmons. (b,d) Maximum size of the 2D cluster state that can be created using (b) FF and (d) TF transmons while maintaining a fidelity of at least 0.8. The parameters used for this calculation are in Table~\ref{tab:2d_gates}.}
    \label{fig:2d_fidelity}
\end{figure*}

To answer the question of which type of transmon can generate better 2D cluster states, we compare the fidelities of $2 \times n$ cluster states generated by either FF or TF transmons. For FF transmon qubits (which were used in Ref.~\cite{Besse2020}), we fix $F_{\text{SQ}} = 0.9995$ and $F_{\text{CR}} = 0.991$ and study the performance as a function of the CNOT fidelity and idling fidelity $F_{\text{idle}}$, which is controlled by the photon generation window $\tau$ [Eq.~\eqref{eq:decay_decoh_loss}]. We always require the emission window $\tau$ to be much longer than the lifetime of the emitter qubit when the coupling is turned on. For TF transmons, we tune the CZ gate fidelity $F_{\text{CZ}}$ and the emission window $\tau$, while fixing $F_{\text{SQ}} = 0.9995$. In Fig.~\ref{fig:2d_fidelity}a,c, we show the infidelity of generating a single column of the $2 \times n$ cluster state, i.e., $2$ photons, using FF transmons (a) and TF transmons (c), respectively. In Fig.~\ref{fig:2d_fidelity}b,d, we also show the maximum size of the 2D cluster state that can be created while maintaining a fidelity of at least $0.8$. The performance of both types of transmon qubits is very similar. In both cases, we can generate $16$-photon cluster states while maintaining a state fidelity of $0.8$.

So which type of transmon is better? Comparing Figs.~\ref{fig:2d_fidelity}a and c, we notice that FF transmon qubits perform better compared to TF transmons when the spontaneous emission time is similar in the two cases. To quantify this improvement, we can compare Eqs.~\eqref{eq:fix_2D_all} and \eqref{eq:tunable_2D_all} under the assumption that the single- and two-qubit gates are of the same order in the two cases. The ratio of the fidelities is then
\begin{equation}
    \mathcal{F} \overset{\mathrm{def}}{=} F_{\text{SQ}}^{-3/k} \frac{ F_{\text{idle}}^{(\text{FF})} (\tau^{(\text{FF})})}{F_{\text{idle}}^{(\text{TF})}(\tau^{(\text{TF})})}.
    \label{eq:2d_cluster_criteria}
\end{equation}
If $\mathcal{F} > 1$, the FF transmon scheme wins, while otherwise TF transmons work better. When the emission window is long, the higher loss and decoherence errors in the TF case dominate, making the FF transmons preferable. On the other hand, when the emission window is short, the loss and decoherence errors are similar for both types of transmons, $F_{\text{idle}}^{(\text{FF})}/F_{\text{idle}}^{(\text{TF})} \sim 1$, while for $k = 2$, $\mathcal{F} \sim F_{\text{SQ}}^{-3/2} >1$, meaning the FF transmon qubit scheme again performs better, this time because it requires fewer SQGs compared to the TF scheme. In the parameter regime shown in Fig.~\ref{fig:2d_fidelity}, the ratio $\mathcal{F}$ in Eq.~\eqref{eq:2d_cluster_criteria} is close to unity, and both types of transmons yield very similar results. Note that very small emission windows ($\tau<200$~ns) can only be achieved for TF transmons, in which case $F_{\text{idle}}^{\text{(TF)}}$ can be greater than $F_{\text{idle}}^{\text{(FF)}}$, and better performance can be achieved using TF transmons provided $F_{\text{SQ}}$ is sufficiently close to unity.

\section{Generating microwave tree graph states and repeater graph states} \label{sec:rgs}

In this section, we show how to generate microwave tree graph states and repeater graph states (RGS), which are useful for robust quantum communication using microwave photons~\cite{Varnava2006, Pant2017, Azuma2015, Buterakos2017, Li2019}.
We start by reviewing existing protocols for generating optical tree graph states and RGSs using quantum emitters. We then propose superconducting qubit devices for generating microwave tree graph states and RGSs and estimate the fidelities of the generated states.

\begin{figure}[h]
    \centering
    \includegraphics[width = 0.42 \textwidth]{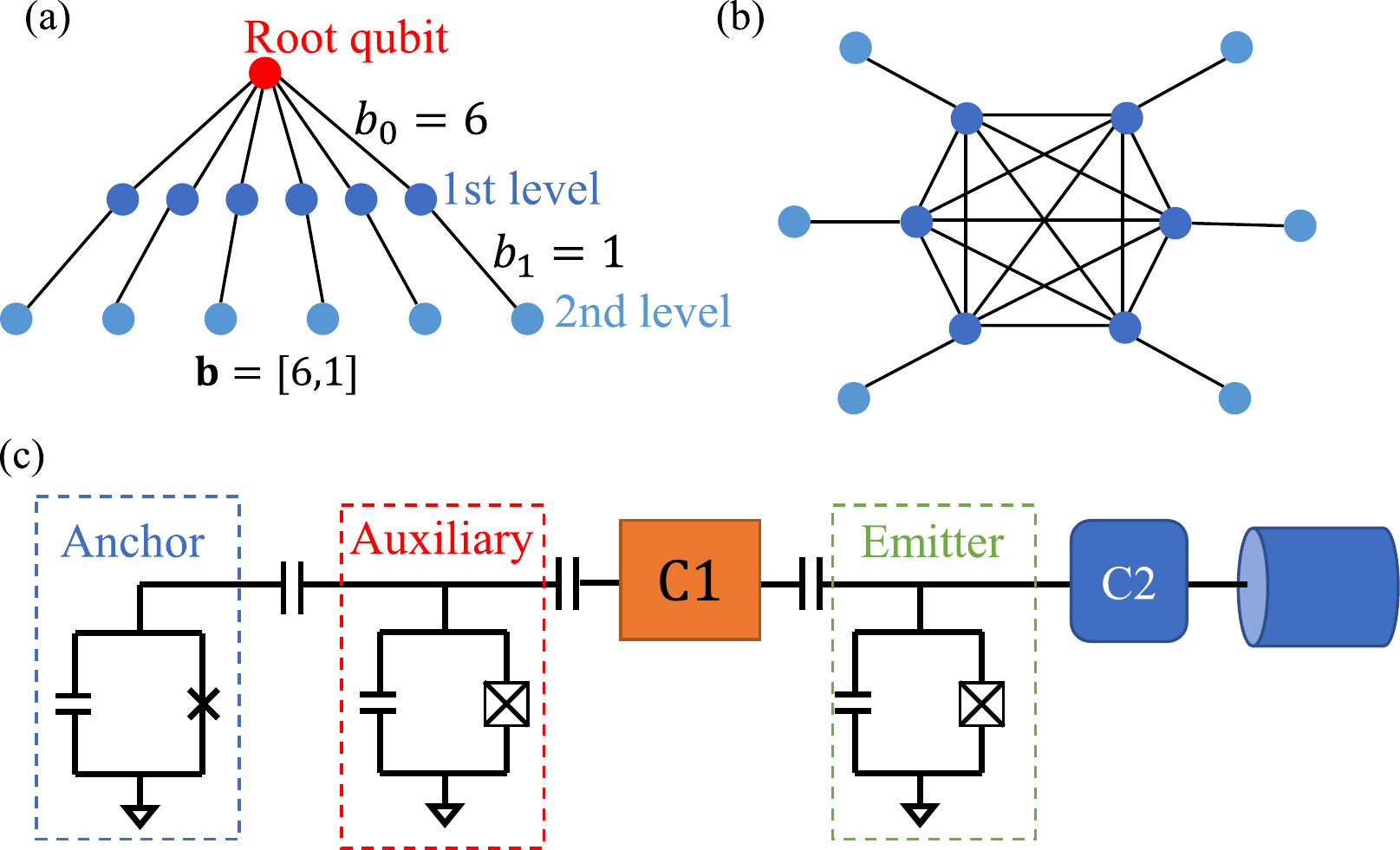}
    \caption{(a) Tree graph state with $6$ arms, each of which has 2 photonic qubits. The branching vector for this graph is $\mathbf{b} = [6,1]$. This state can be converted into the RGS shown in (b) by a Pauli-$Y$ measurement on the root qubit, up to local unitaries on the first-level qubits. (c) The proposed superconducting circuit for generating RGSs like that shown in (b).}
    \label{fig:rgs_circuit}
\end{figure}

Quantum repeaters play an important role in quantum communication through lossy quantum channels. Repeater schemes based on matter qubits require these qubits to have coherence times that are long compared to the time scale for building entanglement~\cite{Duan2001, Childress2006, Hartmann2007, Collins2007, Sangouard2011, Vinay2017, Rozpedek2019, Bhaskar2020}. 
In Ref.~\cite{Azuma2015}, Azuma \textit{et al.} proposed to use all-photonic repeaters based on RGSs to avoid coherence time requirements and other issues associated with memory-based repeaters. In Fig.~\ref{fig:rgs_circuit}b, we show the graph corresponding to a 12-photon RGS.

On the other hand, tree graph states (see for example, Fig.~\ref{fig:rgs_circuit}a) are important for making quantum communication and measurement-based quantum computation robust against photon loss~\cite{Varnava2006}. Such graphs are characterized by a vector of branching parameters $\mathbf{b} = [b_0, b_1, ...]$ that indicate how many branches extend out of each vertex at a given level of the tree. For example, $b_0$ is the number of branches extending out of the 0th-level ``root" vertex at the top of the tree, while $b_1$ is the number of branches emerging from each of the vertices at level 1 (the ``children" of the root vertex). Fig.~\ref{fig:rgs_circuit}a shows an example with 12 photons and branching vector $\mathbf{b} = [6,1]$.
Tree graphs can be combined with other graph states to create states of logically encoded qubits that are robust against photon loss~\cite{Varnava2006, Pant2017}. This approach allows measurement outcomes of lost photons to be inferred from outcomes of measurements on neighboring photons in the tree graph. Inserting tree graphs into RGSs yields all-photonic repeaters that are robust to photon loss ~\cite{Varnava2006, Buterakos2017, Hilaire2021}.

Refs.~\cite{Azuma2015, Buterakos2017} pointed out that RGSs can be obtained from certain tree graph states using only single-qubit measurements. Specifically, a tree graph with branching vector $[b_0,1]$ can be converted into an RGS with $2b_0$ photons by applying a Pauli-$Y$ measurement on the root qubit and SQGs on the first-level qubits. Thus, the state in Fig.~\ref{fig:rgs_circuit}b can be obtained from the state in Fig.~\ref{fig:rgs_circuit}a in this way.

The task of generating RGSs shown in Fig.~\ref{fig:rgs_circuit}b then amounts to generating tree graph states with branching parameters $[6,1]$. Refs.~\cite{Buterakos2017,Russo2019} showed that this can be done in the optical domain using only two matter qubits with the right level structure and selection rules (such as quantum dot spins); one qubit serves as the emitter and the other one as an ``anchor" qubit. In this approach, each 2-photon arm of the tree is generated by the emitter and then connected to the anchor (the root qubit in Fig.~\ref{fig:rgs_circuit}a) by (1) entangling the emitter and anchor qubits, (2) pumping the emitter twice to generate two entangled photons, applying a Hadamard on the emitter before each pumping, and (3) measuring the emitter qubit in the Pauli-$Z$ basis. Repeating this process six times, the $[6,1]$ tree graph state is generated and ready for the final Pauli-$Y$ measurement on the anchor qubit to yield the RGS. This measurement must be followed by single-qubit $Z$ rotations on all photons that were connected to the anchor in order to restore the remaining state to a proper graph state.

This approach can be generalized to create depth-2 tree graph states with any branching vector $[b_0, b_1]$ by allowing the emitter qubits to emit $b_1+1$ photons in step (2). It is also straightforward to generalize the scheme to create a tree graph state with $n$ levels ($\mathbf{b} = [b_0, b_1, ..., b_{n-1}]$) using $n-1$ matter qubits \cite{Hilaire2021}. Such a tree graph state can be generated level-by-level: (1) using two emitters to generate a tree graph state with branching number $[b_{n-2},b_{n-1}]$, (2) connecting the root emitter of this sub-tree to another emitter and pumping the sub-tree root emitter once followed by measuring the emitter in the Pauli-$X$ basis, (3) repeating (1) and (2) $b_{n-3}$ times to create a tree graph state with branching number $[b_{n-3},b_{n-2},b_{n-1}]$ rooted on the new emitter, and (4) repeating the steps above until the complete tree graph state is prepared. In the following, we show how to implement the essential steps in generating depth-2 microwave tree graph states and RGSs and estimate the state fidelity.

To generate tree graph states and RGSs in the microwave regime, we propose using the superconducting circuit shown in Fig.~\ref{fig:rgs_circuit}c.
This circuit consists of a total of three transmon qubits: one emitter, one auxiliary, and one anchor. The anchor plays the same role as in the optical generation scheme. We capacitively couple the anchor qubit and the auxiliary qubit in the PGU together, which allows the application of CZ gates between them. Even though it is necessary to apply SQGs on some of the photonic qubits as discussed above in context of optical schemes, we can still use the Fock-basis encoding. This is because the Pauli-$Z$ rotations on photonic qubits can be virtually applied by adjusting the lab frame.
We stress that microwave tree graph states and RGSs based on the other three photonic encodings reviewed in Sec.~\ref{sec:toolbox} can also be generated in a manner similar to what is described in this section. However, as there are two emission windows for each photonic qubit in these other three cases, the overall state fidelity will be limited compared to using the Fock-basis encoding.

Using the generation circuit in Fig.~\ref{fig:rgs_circuit}, a microwave $[b_0, b_1]$ tree graph state can be generated with the following steps:
\begin{enumerate}
    \item Prepare the anchor and auxiliary qubits in the $\ket{+} = ( \ket{0} + \ket{1} ) /\sqrt{2}$ state.
    \item Apply a CZ gate between the auxiliary and anchor qubits.
    \item Apply a CNOT gate between the auxiliary and emitter qubits, followed by a Hadamard gate on the emitter qubit, and then photon emission from the emitter qubit to generate a Fock-basis encoded photonic qubit.
    \item Repeat the previous step $b_0+1$ times, but omitting the emitter Hadamard gate before the last photon emission.
    \item Perform Pauli-$X$ measurements on the auxiliary qubits.
    \item Repeat step 2 to step 5 $b_0$ times.
\end{enumerate}
After generating the $[b_0, b_1]$ tree graph state, applying a Pauli-$Y$ measurement on the anchor qubit turns the tree graph state into an RGS, in which there are $b_0$ fully connected microwave photonic qubits, each connected to $b_1$ second-level photonic qubits. Note that in order for the generated RGS to have high fidelity, the anchor qubit needs to have a very long lifetime and coherence time, much longer than the overall state generation process (repeating step~2 to step~5 $b_0$ times). However, suppose the time window for generating a photon pulse is $\sim 900$~ns as in Ref.~\cite{Besse2020}, and the target state is the RGS shown in Fig.~\ref{fig:rgs_circuit}b. The total generation time is then $\sim 12 \times 900$~ns, which is comparable to the lifetime and coherence time of transmon qubits. It follows that the idling error caused by the decay and decoherence of the anchor qubit is one of the main factors that degrade the generated RGSs.

In order to reduce the idling error on the anchor qubit, we propose to use more PGUs to generate the photon arms in parallel. After all the photon arms are generated, we apply CZ gates between the anchor qubit and the auxiliary qubits in all PGUs. And then we can measure the auxiliary qubits in the Pauli-$Y$ basis to finalize the tree graph state generation. This parallelized generation scheme is summarized below:
\begin{enumerate}
    \item Prepare the auxiliary qubits in all $b_0$ PGUs in the $\ket{+}$ state.
    \item Use CNOT gates between the auxiliary and the emitter qubit, combined with SQGs, to generate $(b_1 + 1)$ Fock-basis encoded photonic qubits from all PGUs.
    \item Initialize the anchor qubit in the $\ket{+}$ state and apply CZ gates between the anchor qubit and the auxiliary qubits in each PGU.
    \item Measure all the auxiliary qubits in the Pauli-$Y$ basis, and depending on the measurement result, apply a Pauli-$Z$ gate (virtually) on the last photon qubits generated in step~3.
\end{enumerate}
In this way, the idling error can be reduced, as the anchor qubit does not need to wait for all the photonic qubits to be generated. In the following, we estimate the fidelity of the photonic microwave RGSs based on this parallelized generation scheme. First we consider FF transmons, and we then repeat the analysis for TF transmons.

\begin{figure}[h]
    \centering
    \includegraphics[width = 0.42 \textwidth]{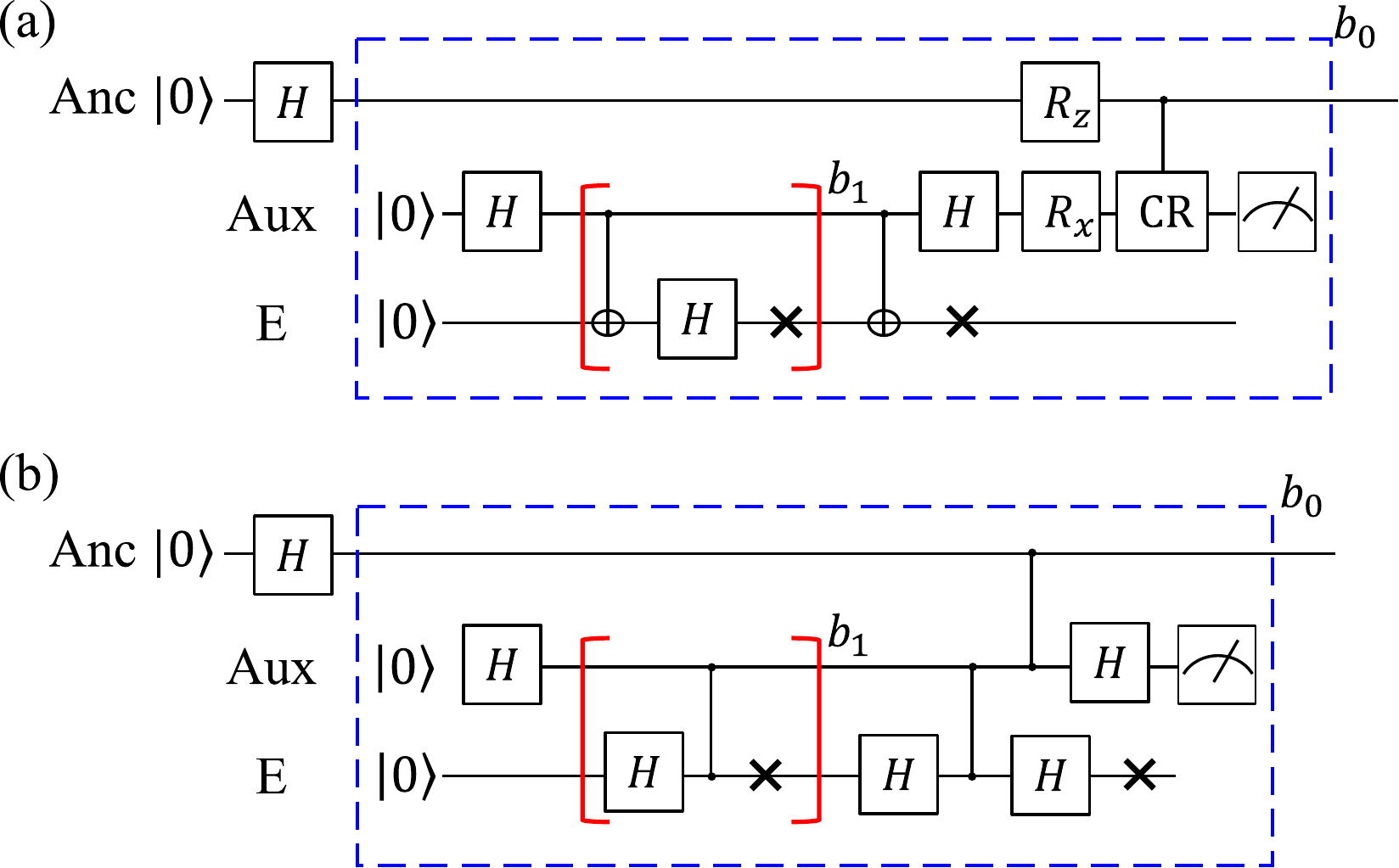}
    \caption{The gate sequence to generate Fock-basis encoded $[b_0, b_1]$ tree graph states rooted on the anchor qubit using (a) FF transmons or (b) TF transmons. The red bracket with super-index $b_1$ means the gates inside the bracket are repeated $b_1$ times. The blue dashed box with super-index $b_0$ means that there are $b_0$ copies of qubits and gates inside the box. The black crossings are microwave photon emissions. Pauli rotations $R_z$ and $R_x$ are rotation by $\pi/2$ by default. `Anc', `Aux' and `E' stand for anchor, auxiliary and emitter qubit, respectively.}
    \label{fig:rgs_gate_seq}
\end{figure}

Consider generating a $[b_0, b_1]$ tree graph state using FF transmon qubits. The gate sequence is shown in Fig.~\ref{fig:rgs_gate_seq}a. Generating a photon arm with $b_1 + 1$ photon qubits that connect to the auxiliary qubit in the PGU requires one Hadamard gate on the auxiliary qubit (for initialization), $(b_1+1)$ CNOT gates, and $b_1$ Hadamard gates on the emitter qubit. When connecting the auxiliary qubit to the anchor qubit, we need to decompose the CZ gate into a CR gate and SQGs. We also combine consecutive SQGs (the Hadamard gate and the $R_x$ rotation) on a transmon qubit into a single SQG. After connecting the auxiliary qubit to the anchor, we measure the auxiliary qubit in the Pauli-$Z$ basis. If the measurement result is $-1$, we virtually apply a Pauli-$Z$ gate to the last emitted photon qubit. The fidelity of a single photon arm of the $[b_0, b_1]$ tree graph state is then
\begin{equation}
    F = F_{\text{SQ}}^{b_1+3} F_{\text{CNOT}}^{b_1+1}  F_{\text{CR}} \left[F_{\text{idle}}(\tau)\right]^{b_1+1} F_{\text{m}},
\end{equation}
where $F_{\text{idle}}(\tau)$ is the idling error on the auxiliary qubit while waiting for a photonic qubit to be emitted, $\tau$ is the corresponding emission time, and $F_{\text{m}}$ is the measurement fidelity on the auxiliary qubit. The fidelity of the total $[b_0, b_1]$ tree graph state is then estimated to be
\begin{equation}
    F_{[b_0, b_1]}^{(\text{FF})} = F_{\text{SQ}}^{N + 2 b_0 + 1} F_{\text{CNOT}}^{N} F_{\text{CR}}^{b_0} \left[F_{\text{idle}}^{(\text{FF})}(\tau)\right]^{N}  F_{\text{m}}^{b_0},
    \label{eq:rgs_fixed}
\end{equation}
where $N = b_0 (b_1 +1)$ is the total number of photonic qubits inside the $[b_0, b_1]$ tree graph state. To generate the microwave RGS, the crucial operation is the Pauli-$Y$ measurement on the anchor qubit, which will decrease the state fidelity by an additional factor of $F_{\text{SQ}}F_{\text{m}}$ relative to Eq.~\eqref{eq:rgs_fixed}.

On the other hand, if we use TF transmons, the CNOT gates between the auxiliary and emitter qubits are decomposed into CZ gates and Hadamard gates on the emitter qubits, while the CZ gates between the anchor and auxiliary qubits inside the PGUs can be directly applied. The gate sequence is shown in Fig.~\ref{fig:rgs_gate_seq}b, where the Hadamard gate on the emitter from decomposing the CNOT gate is cancelled by the Hadamard gate before the photon emission. So the fidelity of the $[b_0, b_1]$ tree graph state is 
\begin{equation}
    F_{[b_0, b_1]}^{(\text{TF})} = F_{\text{SQ}}^{N + 3 b_0 + 1} F_{\text{CZ}}^{N + b_0}   \left[F_{\text{idle}}^{(\text{TF})}(\tau)\right]^{N} F_{\text{m}}^{b_0}.
    \label{eq:rgs_tunable}
\end{equation}
Similarly, the fidelity of the RGS generated from the $[b_0, b_1]$ tree graph state is $F_{\text{SQ}} F_{\text{m}} F_{[b_0, b_1]}^{(\text{TF})}$, where $F_{\text{SQ}}F_{\text{m}}$ accounts for the Pauli-$Y$ measurement on the anchor qubit.

Comparing the results for FF and TF transmons, Eqs.~\eqref{eq:rgs_fixed} and~\eqref{eq:rgs_tunable}, we see that TF transmons require more SQGs on the emitter qubits. Assuming the single- and two-qubit gate fidelities are comparable in the two cases, we define the quantity
\begin{equation}
    \mathcal{F}_{[b_0, b_1]} \overset{\mathrm{def}}{=} F_{SQ}^{-\frac{1}{b_1+1}} \left[ \frac{F_{\text{idle}}^{(\text{FF})} (\tau^{(\text{FF})})}{F_{\text{idle}}^{(\text{TF})}(\tau^{(\text{TF})})} \right],
\end{equation}
where $\tau^{(\text{FF})}$ and $\tau^{(\text{TF})}$ are the emission time windows for FF and TF transmons, respectively. When $\mathcal{F}_{[b_0, b_1]} > 1$, FF transmons yield higher-fidelity RGSs, while for $\mathcal{F}_{[b_0, b_1]}<1$, TF transmons perform better.

\begin{figure}[htbp]
    \centering
    \subfloat[Fixed-frequency transmon qubits]{\includegraphics[width = 0.475 \textwidth]{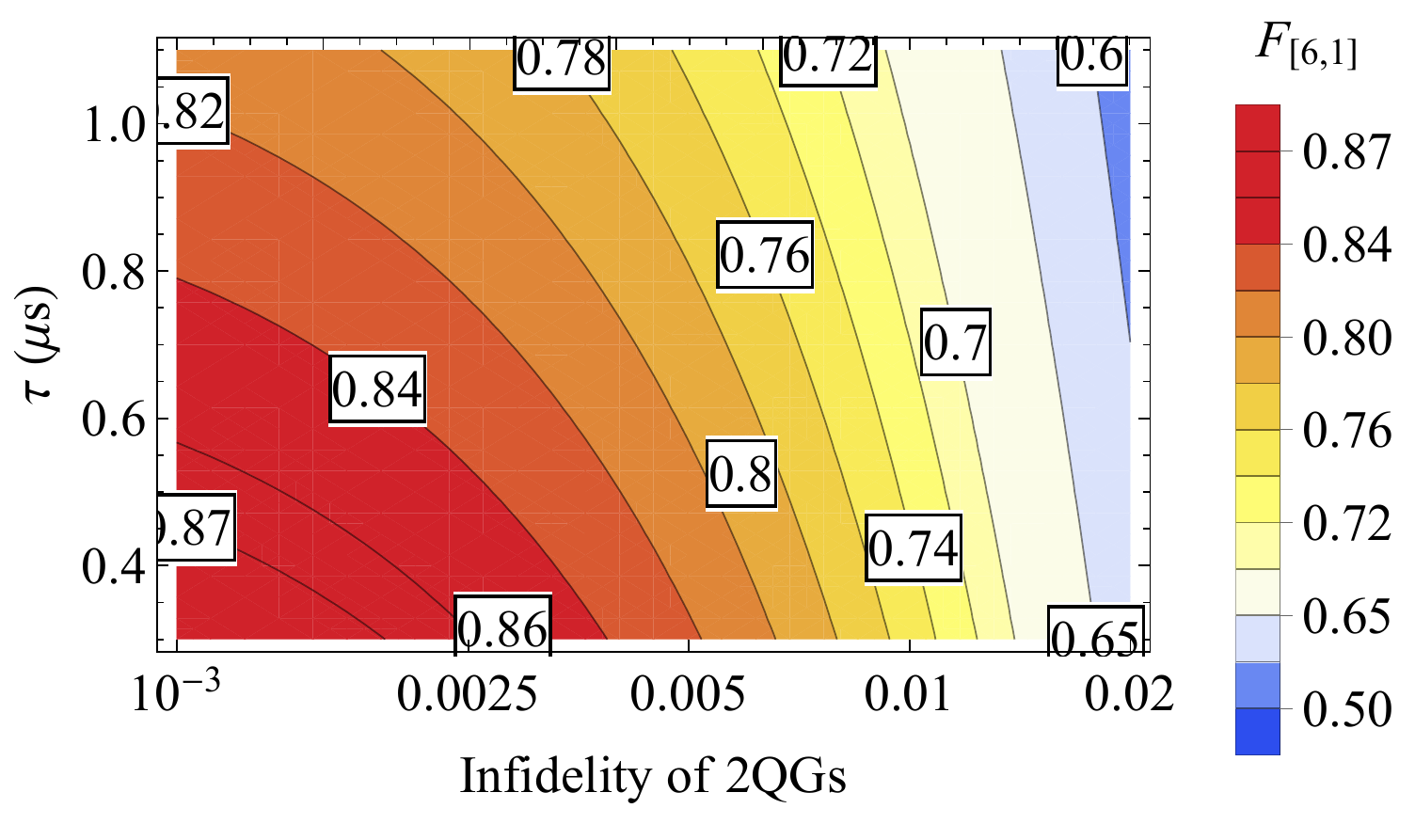}}\\
    \subfloat[Tunable-frequency transmon qubits]{\includegraphics[width = 0.475 \textwidth]{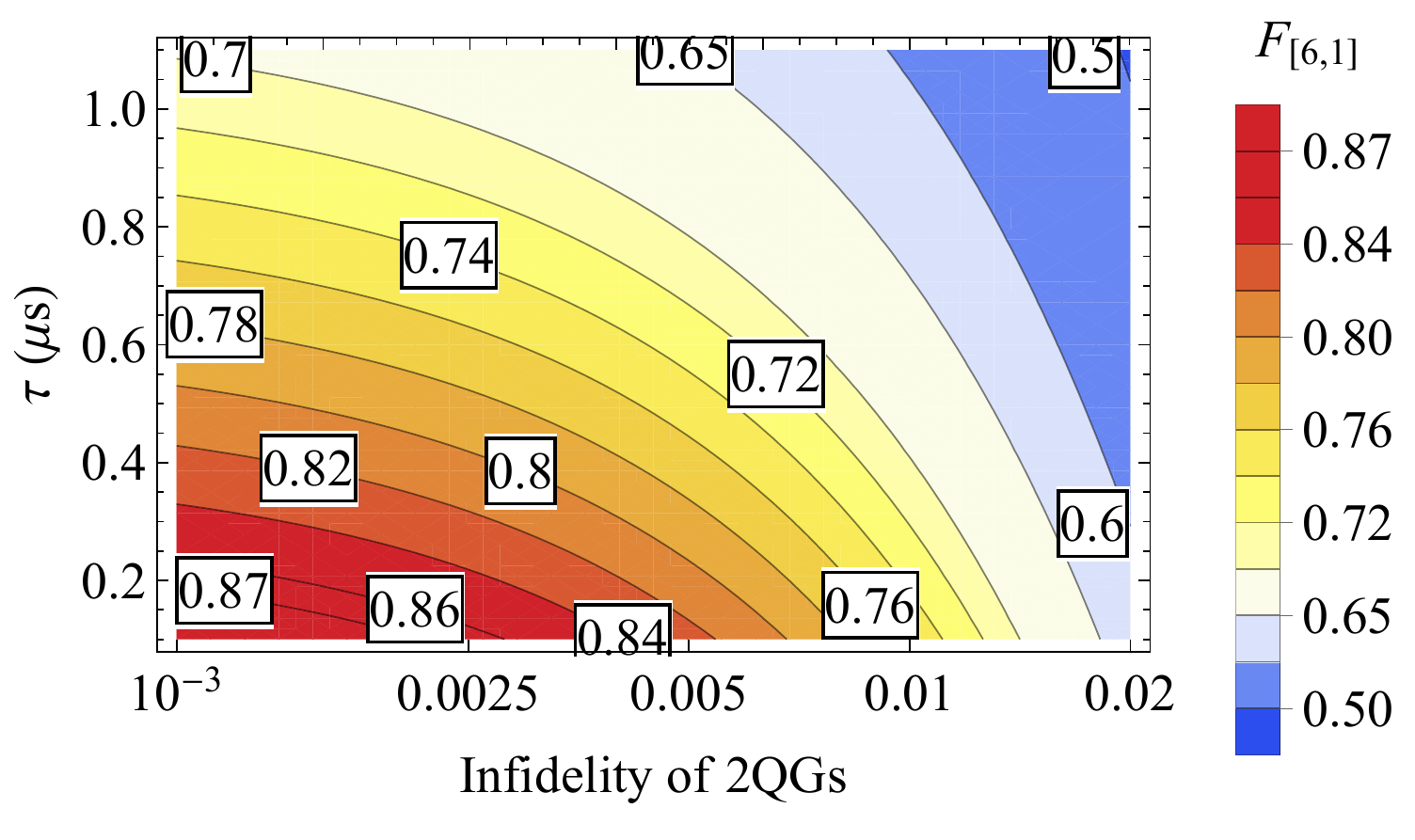}}
    \caption{Fidelity of the $[6,1]$ tree graph state shown in Fig.~\ref{fig:rgs_circuit}a and generated by (a) FF transmon qubits and (b) TF transmon qubits. The fidelities are obtained from Eqs.~\eqref{eq:rgs_fixed} and \eqref{eq:rgs_tunable}. We set $F_{\text{m}} = 0.99$, while remaining parameter values are given in Tables~\ref{tab:2d_gates}. We tune the microwave qubit generation time $\tau$ and the fidelity of the two-qubit gates (2QG) between transmons. We tune the fidelities of CNOT and CR gates for FF transmons, while we tune the fidelity of CZ gates for TF transmons.}
    \label{fig:rgs_fidelity}
\end{figure}

Figs.~\ref{fig:rgs_fidelity}a and~\ref{fig:rgs_fidelity}b show the fidelity of the $[6,1]$ tree graph state depicted in Fig.~\ref{fig:rgs_circuit}a and generated from FF and TF transmons, respectively. In both cases, we examine the performance as a function of emission time $\tau$ and two-qubit gate infidelity (CNOT and CR in the FF case, CZ in the TF case). The values of other parameters used in the calculation are summarized in Table~\ref{tab:2d_gates}, and we set the measurement fidelity to $F_{\text{m}} = 0.99$. 

In the case of FF transmons, Fig.~\ref{fig:rgs_fidelity}a shows that when the infidelity of the two-qubit gate is high ($\Delta F_{\text{CNOT}} \sim 0.01$ to $0.02$), the fidelity of the RGS does not significantly depend on the emission time $\tau$. This is because the infidelity of the state is dominated by the flawed two-qubit gate operation in this regime. When the CNOT gate improves to $F_{\text{CNOT}} \sim 0.998$ (infidelity $\Delta F_{\text{CNOT}} = 0.002$), reducing the emission time $\tau$ becomes beneficial for improving the fidelity. Note that the state-of-the-art infidelity of the CNOT gate is on the order of $0.01$~\cite{Kandala2021}. 
Setting $F_{\text{CNOT}} = 0.99$ and $\tau = 900$~ns, the $[6,1]$ tree graph state fidelity is $F_{[6,1]} \sim 0.71$ 
However, if the two-qubit gate fidelity is improved to $0.995$, the state fidelity increases to $F_{[6,1]} \sim 0.77$, while if instead the emission time is reduced to $\tau = 0.3~\mu$s, the fidelity only reaches $F_{[6,1]} \sim 0.76$. Implementing both improvements yields $F_{[6,1]} \sim 0.82$. If the two-qubit gate on transmons can be improved to $F_{\text{CNOT}} = F_{\text{CNOT}} = 0.998$, and the emission time to $\tau = 0.3~\mu$s, then the state fidelity $F_{[6,1]}$ reaches $0.87$.

We now turn to the results for TF transmons (Fig.~\ref{fig:rgs_fidelity}b). Note that for TF transmons, the relaxation time can be as low as $\sim 1$~ns~\cite{Zhong2019}, making the emission time much shorter than in the FF case. This allows us to consider emission windows as short as $\tau \sim 100$~ns. For CZ gates of fidelity $\sim 0.995$, the loss and decoherence errors on the transmon qubits during the time $\tau$ are important, and decreasing $\tau$ can improve the tree graph state fidelity significantly. Note that $F_\text{CZ} \sim 0.995$ is experimentally achievable~\cite{Barends2014, Sung2021}. With $\tau = 0.9~\mu$s and $F_\text{CZ} = 0.995$, the fidelity of the $[6,1]$ tree graph state is $0.68$. However, if the emission time can be improved to $\tau = 0.1~\mu$s, the fidelity becomes $0.83$. If we further improve the CZ gate fidelity to $F_{\text{CZ}} = 0.998$, the fidelity of the state can be $F_{[6,1]} \sim 0.87$.

\begin{table*}[htbp]
\caption{\label{tab:error_tree} Error analysis of the tree graph state generation schemes. We remove each error source one-at-a-time and calculate how much the fidelity of the $[6,1]$ tree graph state can be improved. The fidelity of the tree graph state we start with corresponds to setting $\tau = 0.3~\mu$s, $F_{\text{CNOT}} = F_{\text{CR}} = 0.995$ for FF transmons, and $\tau = 0.1~\mu$s, $F_{\text{CZ}} = 0.995$ for TF transmons. The other parameters are summarized in Tables~\ref{tab:2d_gates} and we set $F_{\text{m}} = 0.99$. SQG: single-qubit gates, 2QG: two-qubit gates (CNOT, CR and CZ gates). 
}
\begin{ruledtabular}
\begin{tabular}{p{3 cm}|p{1.5 cm}|c c c c}
\raggedright Transmon type &\raggedright $F_{[6,1]}$ & \multicolumn{4}{c}{Remove imperfection} \\
 & & SQG & 2QG & Measurement & Loss and Decoherence\\
\hline
Fixed & $0.823$ & $0.833$ & $0.901$ & $0.874$ & $0.850$ \\
Tunable & $0.827$ & $0.839$ & $0.905$ & $0.878$ & $0.847$
\end{tabular}
\end{ruledtabular}
\end{table*}

Finally, we comment on how to improve the state fidelity and we analyze the main error source in the proposed generation schemes. In Table~\ref{tab:error_tree}, we calculate the fidelity of the $[6,1]$ tree graph state while we remove the imperfections one-at-a-time, and compare it with the state fidelity with all the imperfections. We consider removing the imperfections from (1) the SQGs on transmon qubits, (2) the two-qubit gates on transmon qubits, (3) the measurements on the auxiliary qubits, and (4) the loss and decoherence errors on the transmon qubits during the photon emission windows. In both FF and TF transmon schemes, the most detrimental errors (given current experimental parameters) are from the two-qubit gates between transmons.

\section{Generating logically encoded photonic qubits} \label{sec:logical}

In Sections~\ref{sec:2D_cluster} and~\ref{sec:rgs}, we considered how to generate 2D cluster states and RGSs using transmon qubits. In order to make these states more robust to photon loss and other errors, the individual microwave photonic qubits inside the graph state can be replaced by logically encoded qubits. 

The microwave photon emission from a transmon qubit can be modeled as a SWAP gate between the transmon and the emitted photonic qubit, provided we use state $\ket{0}$ to describe the photon before it is emitted. If the emitter qubits are prepared in a logically encoded state, then following emission, the generated photonic qubits will be in the same logically encoded state, while all the emitter qubits will relax back to their ground state $\ket{0}$. More generally, if the emitter qubits are entangled with other qubits, the entanglement will be transferred to the generated photonic qubits after the emission process. These features allow our proposed graph state generation circuits to be easily generalized to generate logically encoded microwave graph states by using more emitter qubits.

\begin{figure}[h]
    \centering
    \includegraphics[width = 3.4 in]{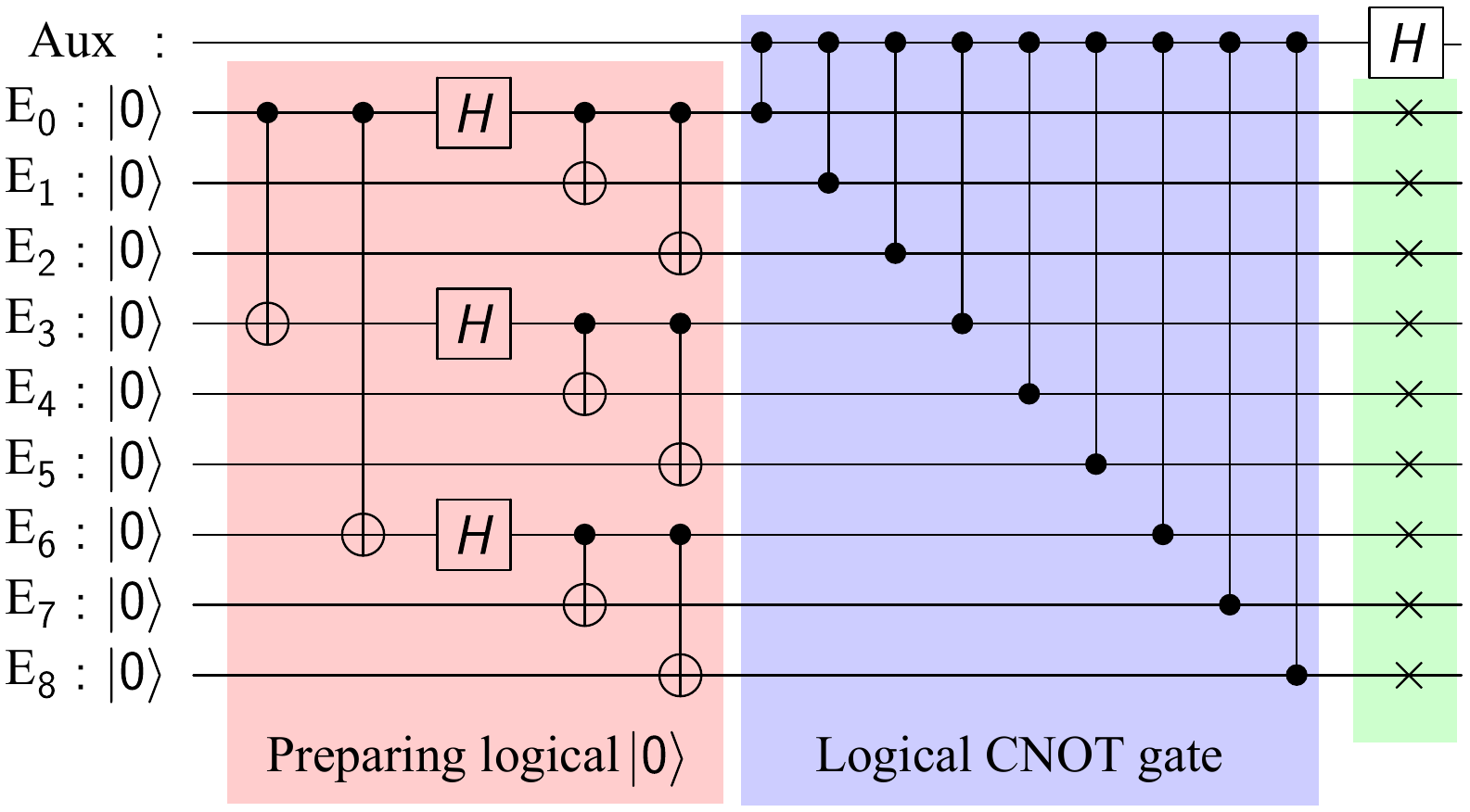}
    \caption{The gate sequence for generating a 9-qubit Shor code encoded logical microwave photonic qubit using transmon qubits. The red-shaded gates are for preparing the logical $\ket{0}$ state for the 9 emitter qubits~\cite{Nielsen2010book}. The blue shaded gates are to perform the logical CNOT gate between the auxiliary qubit and the logical encoded emitter qubits. The green shaded crossings show the microwave photons emission process of the emitter qubits.}
    \label{fig:9_qubit_Shor}
\end{figure}

Based on these observations, we see that we can generate logically encoded photonic graph states if we replace each emitter qubit by several transmon qubits that are initialized in the logical $\ket{0}$ state before the logical CNOT gates between the auxiliary qubit(s) and the emitter qubits are applied. Here we consider the 9-qubit Shor code as an explicit example, although the same generation scheme can also be applied to other error correcting codes, e.g, 7-qubit Steane code, parity code, etc. We consider using 9 transmon qubits as the quantum emitter qubits. Although we can also use a logically encoded auxiliary qubit to further improve the robustness, for demonstration purposes, we consider using a single transmon qubit as an auxiliary qubit in the generation unit. In Fig.~\ref{fig:9_qubit_Shor}, we show the gate sequence to generate a single microwave logically encoded qubit inside a 1D cluster state. The gates that are shaded in red are for preparing the 9 emitter qubits into the logical $\ket{0}$ state of the 9-qubit Shor code. Note that for the 9-qubit Shor code, the logical Pauli-$X$ operation is 
\begin{equation}
    X_{L} = \prod_{i=1}^9 Z_i,
\end{equation}
where $Z_i$ is the Pauli-$Z$ operation on the $i$-th physical qubit. Therefore, the CNOT operation between the single auxiliary transmon qubit and the logically encoded nine emitter qubits can be done using CZ gates as shown in the blue shaded gates in Fig.~\ref{fig:9_qubit_Shor}. After the photon emission from the emitter qubits (crossings in Fig.~\ref{fig:9_qubit_Shor}, shaded in green), a new logically encoded microwave qubit is generated, which contains 9 physical photonic qubits. After applying a Hadamard gate operation to the auxiliary qubit, we finish adding a new logically encoded photonic qubit into the 1D cluster state. This scheme can be easily generalized to produce 2D cluster states, tree graph states, and photonic quantum emitter states with encoded photonic qubits.

\section{Summary and outlook} \label{sec:summary}

In summary, we presented circuits and protocols for generating several important classes of microwave photonic graph states using transmon qubits. In particular, we focused on 2D cluster states, which are resources for measurement-based quantum computing, and repeater graph states, which are useful for robust quantum communication. In each case, we proposed superconducting circuits that can generate these states using either fixed-frequency or tunable-frequency transmon qubits. We also considered four types of photonic qubit encodings and argued that the Fock-basis encoding is the optimal choice for each generation scheme, provided photon loss is not significant. We showed that with current technology, a 2D cluster state with $16$ Fock-basis-encoded photonic qubits can be generated from a circuit containing four transmon qubits with fidelity $0.8$, while the repeater graph states with $12$ photonic qubits as proposed in Ref.~\cite{Azuma2015} can be generated with fidelity $\sim 0.86$. We found that the primary source of infidelity in both cases is the two-qubit gate fidelity.
In addition, we showed that similar circuits and protocols can be used to generate logically encoded qubits that are robust against photon loss. Our results can be used as a guide for how to further improve the generation of microwave graph states from superconducting circuits to achieve high-fidelity quantum communication and transduction between remote superconducting qubit processors.

\section*{Acknowledgement}
S.E.E. acknowledges support from the DOE Office of Science, National Quantum Information Science Research Centers, Co-design Center for Quantum Advantage (C2QA), contract number DESC0012704. This research was in part also supported by the Commonwealth Cyber Initiative (CCI). E.B. acknowledges support from the NSF (grant no. 1561242).

\appendix
\section{Generating encoded photonic qubits using transmon qubits} \label{appsec:photon_qubit_gene}

In this section of the appendix, we provide the explicit gate sequences for generating the different types of encoded photonic qubits using the photon generation units (PGU) presented in the main text. Specifically, we consider three types of encodings: time-bin, frequency-bin, and two-rail encoding. We aim to generate a photonic qubit whose state is entangled with the auxiliary qubit inside the PGU. The generation process can be viewed as a CNOT operation between the auxiliary qubit and the photonic qubit that is virtually prepared in state $\ket{0}$.

\begin{figure}[h]
    \centering
    \includegraphics[width = 0.4 \textwidth]{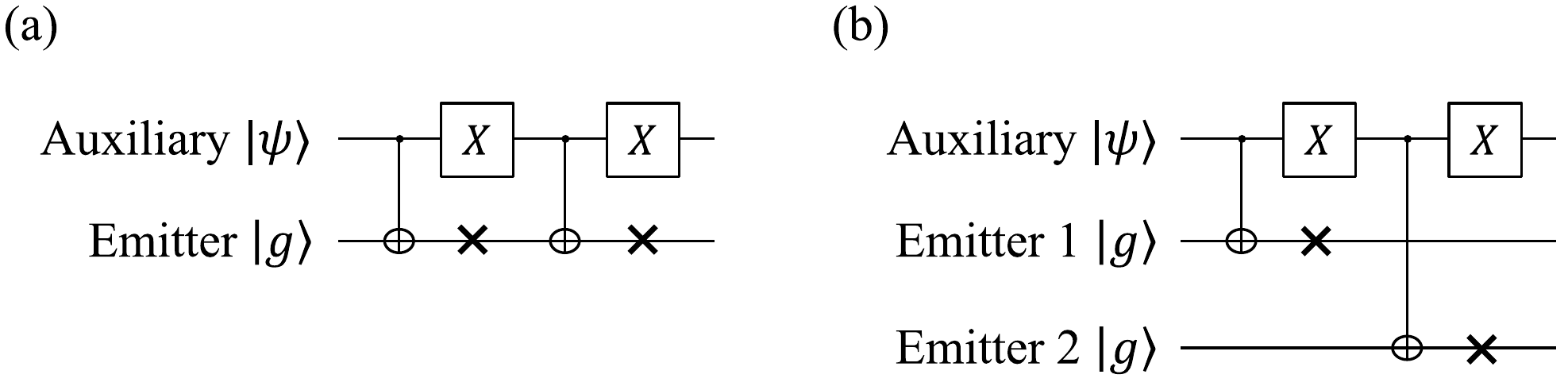}
    \caption{Gate sequences for generating individual photonic qubits. In (a), we show the gate sequence for generating a time-bin encoded photonic qubit, while in (b), the gate sequence shown can be used to generate a frequency-bin or two-rail encoded photonic qubit. The crosses denote photon emission to the corresponding transmission line mode.}
    \label{fig:photon_generation}
\end{figure}

Time-bin encoded photonic qubits use the photon pulses in early or late time bins to encode quantum information. The PGU needs one emitter qubit to generate a time-bin encoded photonic qubit. The generation gate sequence is in Fig.~\ref{fig:photon_generation}a. Because the photon pulses in the early and late time-bins are generated by the same emitter, and the two time-bins need to be well separated in the time domain, the time for generating a single time-bin encoded photon qubit is $\tau_{\text{tb}} = 2 \tau_{\text{CNOT}} + 2 \tau_0$, where $\tau_{\text{CNOT}}$ is the time for applying a CNOT gate, $\tau_0$ is the time to emit a single photon.

Frequency-bin encoded photonic qubits are based on photon pulses with different possible frequencies. In this case, the PGU needs to have two emitter qubits with different transition frequencies. Both emitter qubits in the PGU couple to the same waveguide (transmission line) mode. The gate sequence to generate a frequency-bin encoded photonic qubit is shown in Fig.~\ref{fig:photon_generation}b. As the single-qubit gates and the CNOT gate usually take much less time than the photon emission window $\tau_0$, parallelization is possible. The time for generating a frequency-bin encoded qubit is $\tau_{\text{fb}} = 2 \tau_{\text{CNOT}} + \tau_{\text{X}} + \tau_0$, where $\tau_{\text{X}}$ is the time for a Pauli-$X$ gate on the auxiliary qubit. 

Two-rail encoded photonic qubits use two spatial modes to encode quantum information. They can be generated using two emitter qubits. These two emitter qubits connect to two different transmission lines (waveguides). Similar to the frequency-bin encoded photonic qubit, the two-rail encoded qubits can also be generated by the gate sequence shown in Fig.~\ref{fig:photon_generation}b. Thus, generating a two-rail encoded microwave photonic qubit also takes $\tau_{\text{tr}} = 2 \tau_{\text{CNOT}} + \tau_{\text{X}} + \tau_0$. 

\section{Gate sequences to generate 2D cluster states using fixed- and tunable-frequency transmons} \label{appsec:gate_2D_cluster}

In this section of the appendix, we show the explicate gate sequence for generating a column of photonic qubits inside the $2 \times n$ 2D cluster state using two PGUs. We then discuss the number of gates needed to generate a column of photonic qubits using $k$ PGUs instead. 

\begin{figure}[htbp]
    \centering
    \includegraphics[width = 0.43 \textwidth]{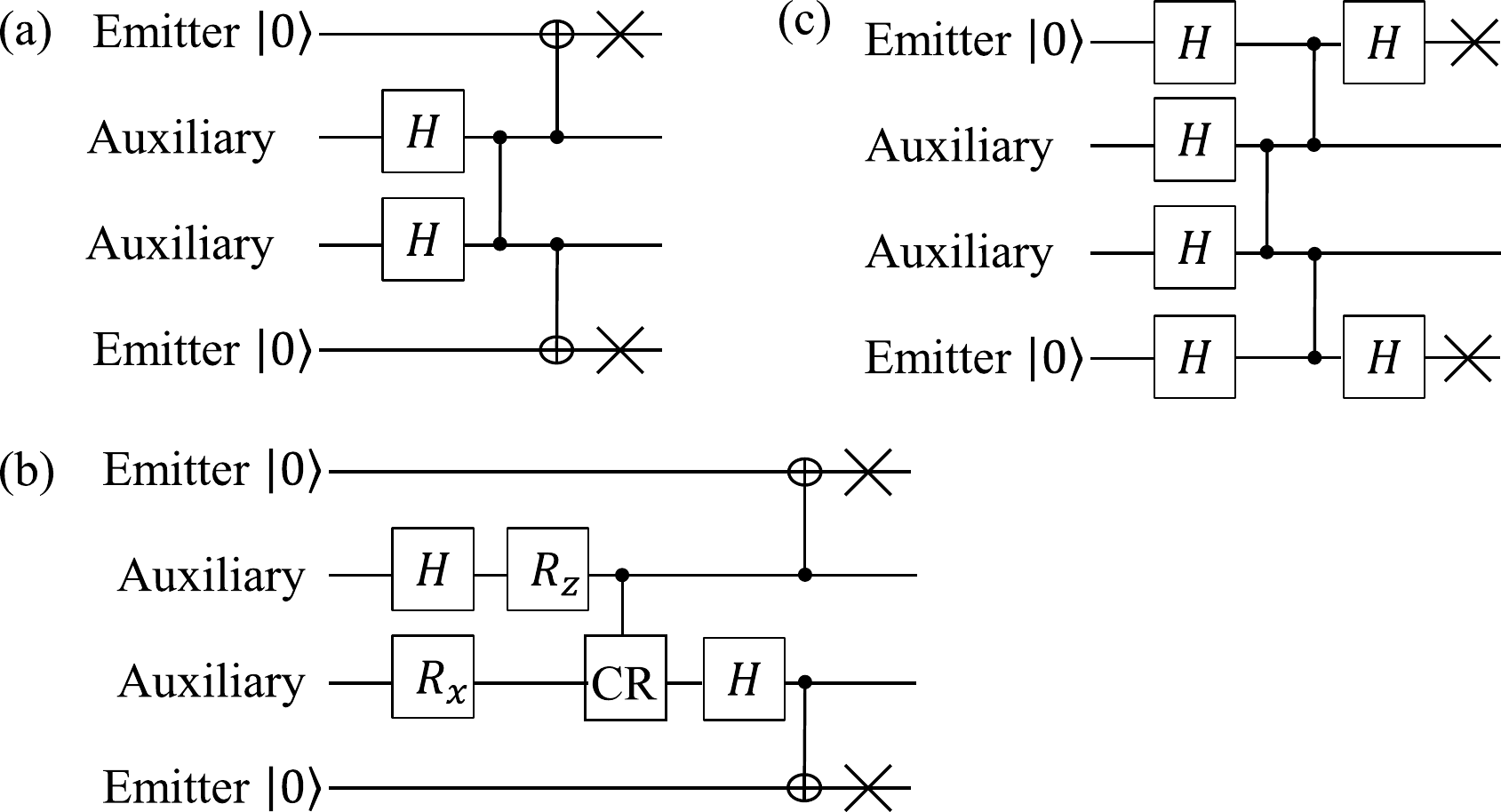}
    \caption{Gate sequences for generating a single column of Fock-basis encoded photonic qubits in a $2 \times n$ 2D cluster state. In (a), we consider using CZ and CNOT gates between transmon qubits. In (b), we decompose the CZ gate in terms of the CR gate, which is one of the native gates of FF transmon qubits. In (c), we decompose the CNOT gate into CZ gates, which are the native gates of TF transmon qubits. The crosses denote photon emission from the corresponding transmon qubit to the connected transmission line.}
    \label{fig:2d_gate_seq}
\end{figure}

In Fig.~\ref{fig:2d_gate_seq}a, we show gate sequences for generating a single column of photonic qubits using two PGUs. We assume that we can use CNOT and CZ gates between the transmon qubits. When we use FF transmon qubits, the CZ gate is not one of the native gates, so we need to decompose it using the CR gate,
\begin{equation}
    U_{\text{CR}} = \frac{1}{\sqrt{2}}\left(
    \begin{array}{cccc}
        1 & i &   &   \\
        i & 1 &   &   \\
          &   & 1 & -i \\
          &   &-i & 1
    \end{array}
    \right),
\end{equation}
and single-qubit gates (SQGs). Note that the CNOT gate can be decomposed using CR gates and Pauli rotations as~\cite{Krantz2019}
\begin{equation}
    \text{C}^{(1)}\text{NOT}^{(2)} = e^{i \pi/4} U_{CR} R_{z}^{(1)}(\pi/2) R_{x}^{(2)}(\pi/2),
\end{equation}
where the superscripts are the qubit index, $R_x$ and $R_z$ are SQGs along the $X$ and $Z$ directions, $\text{C}^{(1)}\text{NOT}^{(2)}$ means a CNOT operation between qubits $1$ and $2$ with $1$ the control and $2$ the target. This yields the gate sequence shown in Fig.~\ref{fig:2d_gate_seq}b. On the other hand, the CNOT gate is not a native two-qubit gate of TF transmon qubits. We decompose CNOT gates into CZ gates with Hadamard gates. The gate sequence for TF transmon qubits is shown in Fig.~\ref{fig:2d_gate_seq}c.

\begin{figure}[h]
    \centering
    \includegraphics[width = 0.4 \textwidth]{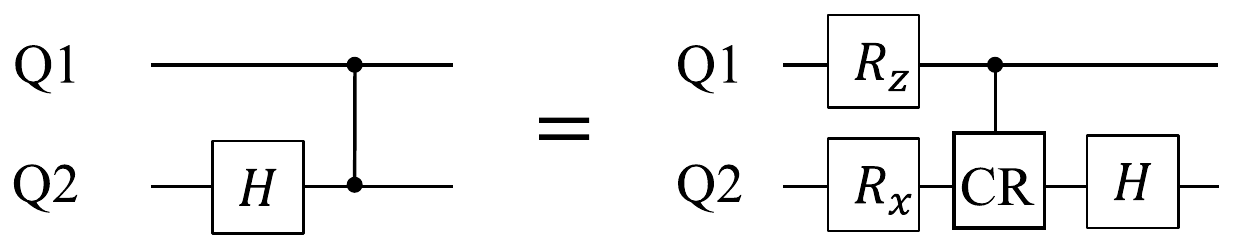}
    \caption{Decomposing a CZ gate in terms of a CR gate and SQGs.}
    \label{fig:block_equiv}
\end{figure}

Next, we consider how the number of gates changes when we instead use $k$ PGUs to generate a column of entangled photonic qubits. For FF transmon qubits, when translating the CZ gates into CR gates, we notice the relation shown in Fig.~\ref{fig:block_equiv}. In the process of generating a column of $k$ photons from $k$ PGUs, there are $k-1$ such blocks. So in each cycle, there are $3(k-1)$ SQGs (note that we combine consecutive SQGs into one, so the first Hadamard gate and $R_x$ rotation are combined into one SQG), $k-1$ CR gates and $k$ CNOT gates, which is reflected in Eq.~\eqref{eq:fix_2D_row}. For TF transmon qubits, the number of gates is easy to count. In each cycle of photon generation, there are $3k$ Hadamard gates, $k-1$ CZ gates between the auxiliary qubits in different PGUs, and $k$ CZ gates between the auxiliary qubits and the emitter qubits.

\section{Estimating the idling fidelity $F_{\text{idle}}$} \label{appsec:idle_F}

In this section, we calculate the idling fidelity $F_{\text{idle}}$ in Eq.~\eqref{eq:decay_decoh_loss}. The idling error is from the decay and decoherence error on the transmon qubit while waiting (idling) for a time window $\tau$. If we treat the decay and decoherence process as a quantum channel, we can calculate its average fidelity to estimate the idling error. Suppose the idling window is $\tau$, the decay and decoherence channel transforms the density matrix of a single qubit as
$$ \rho = \left(
    \begin{array}{cc}
        \rho_{00} & \rho_{01} \\
        \rho_{10} & \rho_{11}
    \end{array}
    \right)
$$
\begin{align}
    \rightarrow \mathcal{E}(\rho) & = \left(
    \begin{array}{cc}
        \rho_{00} + (1 - e^{-\tau/T_1}) \rho_{11} & \rho_{01} e^{-\tau/T_2} \\
        \rho_{10} e^{-\tau/T_2} & e^{-\tau/T_1} \rho_{11} 
    \end{array}
    \right)
\end{align}
where we consider the population relaxation is from the state $\ket{1}$ to the state $\ket{0}$, $T_1$ is the excited state lifetime and $T_2$ is the coherent time. The average fidelity of the decay and decoherence channel is
\begin{equation}
    F_{\text{idle}} = \frac{1}{2} + \frac{1}{6}\left( e^{-\tau/T_1} + 2 e^{-\tau/T_2} \right),
    \label{eq:decay_decoh_loss2}
\end{equation}
after integrating over the Haar measure of single-qubit states. Notice that when the $T_1$ and $T_2$ times of the transmon qubit are given, the idling fidelity $F_{\text{idle}}$ is controlled by the emission window duration $\tau$.

\bibliography{ref}
\end{document}